\documentclass[aps,prb,twocolumn,floatfix]{revtex4}
\usepackage{epsf}

\begin{document}

\title{
Phase transitions in the pseudogap Anderson and Kondo models: \\
Critical dimensions, renormalization group, and local-moment criticality
}

\author{Lars Fritz and Matthias Vojta}
\affiliation{\mbox{Institut f\"ur Theorie der Kondensierten Materie,
Universit\"at Karlsruhe, 76128 Karlsruhe, Germany}}
\date{October 15, 2004}

\begin{abstract}
The pseudogap Kondo problem, describing quantum impurities coupled to fermionic quasiparticles
with a pseudogap density of states, $\rho(\omega)\propto|\omega|^r$, shows a rich
zero-temperature phase diagram, with different screened and free moment phases
and associated transitions.
We analyze both the particle-hole symmetric and asymmetric cases
using renormalization group techniques.
In the vicinity of $r=0$, which plays the role of a lower-critical
dimension, an expansion in the Kondo coupling is appropriate.
In contrast, $r=1$ is the upper-critical dimension in the absence of particle-hole symmetry,
and here insight can be gained using an expansion in the hybridization strength of
the Anderson model.
As a by-product, we show that the particle-hole symmetric strong-coupling
fixed point for $r<1$ is described by a resonant level model, and corresponds
to an intermediate-coupling fixed point in the renormalization group language.
Interestingly, the value $r=1/2$ plays the role of a second lower-critical dimension
in the particle-hole symmetric case, and there we can make progress by a novel
expansion performed around a resonant level model.
The different expansions allow a complete description of all critical fixed
points of the models and can be used to compute a variety of
properties near criticality, describing universal local-moment fluctuations
at these impurity quantum phase transitions.
\end{abstract}
\pacs{75.20.Hr,74.72.-h}

\maketitle


\newcommand{\hybb}  {V_0}               
\newcommand{\hyb}   {v}                 
\newcommand{\epsfb} {\varepsilon_0}     
\newcommand{\epsf}  {\varepsilon}       
\newcommand{\Ub}    {U_0}               
\newcommand{\U}     {u}                 
\newcommand{\Jb}    {J_{\rm K}}         
\newcommand{\lam}   {\lambda_0}         

\section{Introduction}

Non-trivial fixed points and associated phase transitions in
quantum impurity problems have been subject of considerable interest
in recent years,
with applications for impurities in correlated bulk systems, in
transport through nanostructures, and for
strongly correlated lattice models in the framework of dynamical
mean-field theory.
Many of those impurity phase transitions occur in variations
of the well-known Kondo model \cite{hewson} which describes
the screening of localized magnetic moments by metallic conduction
electrons.
A paradigmatic example of an intermediate-coupling impurity fixed point
can be found in the two-channel Kondo effect.

Non-metallic hosts, where the fermionic bath density of states (DOS)
vanishes at the Fermi level, offer a different route to unconventional
impurity physics.
Of particular interest is the Kondo effect in so-called pseudogap
systems \cite{withoff,cassa,tolya2,bulla,GBI,insi,LMA},
where the fermionic bath density of states follows a power law at low energies,
$\rho(\omega) \propto N_0 |\omega|^r$
($r>0$).
Such a behavior arises in semimetals, in certain zero-gap semiconductors,
and in systems with long-range order where
the order parameter has nodes at the Fermi surface, e.g.,
$p$- and $d$-wave superconductors ($r=2$ and 1).
Indeed, in $d$-wave high-$T_c$ superconductors
non-trivial Kondo-like behavior has been observed associated with the magnetic
moments induced by non-magnetic Zn impurities~\cite{bobroff1,bobroff2}.
Note that the limit $r\to\infty$ corresponds to a system with a hard gap.

The pseudogap Kondo problem has attracted substantial attention
during the last decade.
A number of studies \cite{withoff,cassa,tolya2}
employed a slave-boson large-$N$ technique; 
significant progress and insight came from
numerical renormalization group (NRG) calculations \cite{bulla,GBI,insi}
and the local moment approach~\cite{LMA}.
It was found that a zero-temperature phase transition occurs at a critical
Kondo coupling, $J_c$, below which the impurity spin is unscreened
even at lowest temperatures.
Also, the behavior depends sensitively on the presence or absence of
particle-hole (p-h) asymmetry, which can arise, e.g., from a band asymmetry at high
energies or a potential scattering term at the impurity site.
A comprehensive discussion of possible fixed points
and their thermodynamic properties has been given in Ref. \onlinecite{GBI}
based on the NRG approach.

Until recently, analytical knowledge about the critical properties of the
pseudogap Kondo transition was limited.
Previous works employed a weak-coupling renormalization
group (RG) method, based on an expansion in the dimensionless Kondo
coupling $j \!=\! N_0 \Jb$.
It was found that an unstable RG fixed point exists at $j\!=\!r$,
corresponding to a continuous phase transition between the free and screened
moment phases \cite{withoff}.
Thus, the perturbative computation of critical properties within this approach
is restricted to small $r$.
Interestingly, the NRG studies \cite{GBI} showed that the
fixed-point structure changes at $r=r^\ast\approx 0.375$
and also at $r\!=\!\frac{1}{2}$, rendering the relevant case of $r\!=\!1$
inaccessible from weak coupling.
In the p-h symmetric case, for $r\geq\frac{1}{2}$ the phase transition was
found to disappear, and the impurity is always unscreened independent of the
value of $\Jb$.
In contrast, in the asymmetric case the phase transition is present for arbitrary
$r>0$.
Numerical calculations \cite{GBI,insi} indicated that the critical
fluctuations in the p-h asymmetric case change their character at $r\!=\!1$:
whereas for $r\!<\!1$ the exponents take non-trivial $r$-dependent values and obey
hyperscaling, exponents are trivial for $r\!>\!1$ and hyperscaling is violated.
These findings suggest to identify $r\!=\!1$ as upper-critical ``dimension''
of the problem, whereas $r\!=\!0$ plays the role of a lower-critical
``dimension''.
As known, e.g., from the critical theory of magnets \cite{magnets}, the description
of the transitions using perturbative RG requires different theoretical
formulations near the upper-critical and lower-critical dimensions, i.e.,
the $\phi^4$ theory and the non-linear sigma model in the magnetic case.

In this paper, we provide a comprehensive analytical account of the phase
transitions in the pseudogap Anderson and Kondo models,
including the proper theories for the critical ``dimensions''.
This is made possible by working with the Anderson instead of the Kondo
model -- the degrees of freedom of the Anderson model turn out to provide
a natural description of the low-energy physics at the quantum phase transitions
near $r=\frac 1 2$ as well as close to and above $r=1$.
We shall consider epsilon-type expansions in the hybridization, the on-site energy,
and the interaction strength.
Those expansions lead to different theories for the p-h symmetric and
asymmetric cases.
Interestingly, in the pseudogap Kondo model the phase transitions
near the lower-critical and upper-critical dimension are not adiabatically
connected, as the fixed point structure changes both at $r=r^\ast$ and $r=\frac 1 2$.
Thus the present quantum impurity problem has a more complicated
flow structure than the critical theory of magnets, where the $(2+\epsilon)$
and $(4-\epsilon)$ expansions are believed to describe the same critical
fixed point.

In the p-h symmetric case of the pseudogap Kondo problem
the line of non-trivial phase transitions terminates
at {\em two lower-critical dimensions} (!), $r=0$ and $r=\frac 1 2$.
Near $r=\frac{1}{2}$ we find an expansion around a non-interacting
resonant level model, together with perturbative RG, to provide access to
the critical fixed point, with the expansion
being controlled in the small parameter $(\frac{1}{2}-r)$, see Sec.~\ref{sec:sym}.
Interestingly, the weak-coupling expansion for the Kondo model, presented in
Sec.~\ref{sec:weak}, provides a different means to access the same
critical fixed point, but with the small parameter being $r$;
the two expansions can be expected to match.

In the p-h asymmetric case an expansion can be
done in the hybridization around the valence-fluctuation point of the
Anderson model.
Bare perturbation theory is sufficient for all $r>1$;
for $r<1$ a perturbative RG procedure is required to calculate critical properties,
with the expansion being controlled in the small parameter $(1-r)$.
In particular, this identifies $r=1$ as the upper-critical dimension of
the (asymmetric) pseudogap Kondo problem, and consequently observables acquire
logarithmic corrections for $r=1$.
A brief account on the p-h asymmetric case and the expansion around
$r=1$ has been given in a recent paper \cite{MVLF}.
We note that the flow of the asymmetric Anderson model in the metallic
case, $r=0$, was discussed by Haldane\cite{haldane}:
here all initial parameter sets with finite hybridizations flow towards
the strong-coupling (singlet) fixed point.

For all cases listed above, we show that the critical properties of the
Anderson and Kondo models are identical, and we calculate various
observables in renormalized perturbation theory.
To label the fixed points, we will follow the notation of
Ref.~\onlinecite{GBI}.

Before continuing, we emphasize that standard tools
for metallic Kondo models, such as bosonization, Bethe ansatz, and conformal field
theory, are not easily applicable in the present case of a pseudogap density of
states, as the problem cannot be described using linearly dispersing fermions in
one dimension.
Furthermore, integrating out the fermions from the problem, in order to arrive
at an effective statistical mechanics model containing impurity degrees of
freedom only, cannot be performed easily:
the fermionic determinants arising in $(1+r)$ dimensions cannot be simply
evaluated.
This implies that the pseudogap Kondo model does {\em not} map onto
a one-dimensional (e.g. Ising) model with long-ranged
interactions, in contrast to e.g. the spin-boson model \cite{leggett}.
Indeed, the phase transitions in the pseudogap Kondo model and the sub-ohmic
spin-boson model are in different universality classes \cite{BTV}.
Therefore we believe that our combined RG analysis provides a unique
tool for analyzing the pseudogap Kondo problem.


\subsection{Models}
\label{sec:models}

The starting point of our discussion will be the single-impurity
Anderson model with a pseudogap host density of states,
${\cal H} = {\cal H}_{\rm A} + {\cal H}_{\rm b}$:
\begin{eqnarray}
\label{aim}
{\cal H}_{\rm A} &=& \epsfb f_\sigma^\dagger f_\sigma + \Ub n_{f\uparrow} n_{f\downarrow}
  + \hybb \left(f_\sigma^\dagger c_\sigma(0) + {\rm h.c.}\right) \,,
  \\
{\cal H}_{\rm b} &=& \int_{-\Lambda}^{\Lambda} dk\,|k|^r \,
  k c_{k\sigma}^\dagger c_{k\sigma}
\nonumber
\end{eqnarray}
where we have represented the bath, ${\cal H}_{\rm b}$,
by linearly dispersing chiral fermions $c_{k\sigma}$,
summation over repeated spin indices $\sigma$ is implied,
and $c_\sigma(0) = \int d k |k|^r c_{k\sigma}$
is the conduction electron operator at the impurity site.
The spectral density of the $c_\sigma(0)$ fermions
follows the power law $|\omega|^r$ below the
ultra-violet (UV) cutoff $\Lambda$;
details of the density of states at high energies
are irrelevant for the discussion in this paper.
The four possible impurity states will be labelled with
$|\uparrow\rangle$, $|\downarrow\rangle$ for the spin-carrying
states, $|{\rm e}\rangle$ for the empty and $|{\rm d}\rangle$ for the doubly occupied
state.
Provided that the conduction band is p-h symmetric, the above model
obeys p-h symmetry for $\Ub = -2\epsfb$ -- this p-h symmetry can
be considered as SU(2) pseudospin, i.e., the full symmetry of
the model is SU(2)$_{\rm spin}$ $\times$ SU(2)$_{\rm charge}$.
Asymmetry of the high-energy part of the conduction band has the same net
effect as asymmetry of the impurity states; we will always assume that
the low-energy part of the band is asymptotically symmetric, i.e.,
the prefactor of $|\omega|^r$ in the DOS is equal for positive and negative
$\omega$.

The transformation
\begin{eqnarray}
&&f_\sigma \rightarrow f_\sigma^\dagger\,, \nonumber\\
&&c_{k\sigma} \rightarrow c_{k\sigma}^\dagger
\label{ph1}
\end{eqnarray}
converts all particles into holes and vice versa, formally
$\epsfb \rightarrow -(\epsfb+\Ub)$, $\hybb \rightarrow -\hybb$.
Physically, the roles of the states $|\uparrow\rangle$ and $|\downarrow\rangle$
are interchanged, as well as the states $|{\rm e}\rangle$ and $|{\rm d}\rangle$.
It is useful to consider another transformation,
\begin{eqnarray}
&&f_\uparrow \rightarrow f_\uparrow\,,~
f_\downarrow \rightarrow f_\downarrow^\dagger\,, \nonumber\\
&&c_{k\uparrow} \rightarrow c_{k\uparrow}\,,~
c_{k\downarrow} \rightarrow c_{k\downarrow}^\dagger\,,
\label{ph2}
\end{eqnarray}
which transforms $|\uparrow\rangle \leftrightarrow |{\rm d}\rangle$,
$|\downarrow\rangle \leftrightarrow |{\rm e}\rangle$.
Here, the spinful doublet of impurity states is transformed
into the spinless doublet and vice versa, i.e.,
the two SU(2) sectors are interchanged.

In the so-called Kondo limit of the Anderson model
charge fluctuations are frozen out,
and the impurity site is mainly singly occupied.
Via Schrieffer-Wolff transformation one obtains the standard Kondo model,
${\cal H} = {\cal H}_{\rm K} + {\cal H}_{\rm b}$, with
\begin{equation}
\label{km}
{\cal H}_{\rm K} = \Jb {\bf S} \cdot {\bf s}(0)
\label{pgk}
\end{equation}
where the impurity spin $\bf S$ is coupled to the conduction electron
spin at site $0$, $s_\alpha(0) = c_\sigma^\dagger(0) \sigma_{\sigma\sigma'}^\alpha c_{\sigma'}(0) / 2$,
and $\sigma^\alpha$ is the vector of Pauli matrices.
The Kondo coupling is related to the parameters of the Anderson model (\ref{aim})
through:
\begin{equation}
\Jb = 2 \hybb^2 \bigg(\frac{1}{|\epsfb|} + \frac{1}{|\Ub+\epsfb|}\bigg) \,.
\label{j-swo}
\end{equation}
The Kondo limit is reached by taking $\Ub\to\infty$, $\epsfb\to -\infty$,
$\hybb\to\infty$, keeping $\Jb$ fixed.
In the absence of p-h symmetry the Schrieffer-Wolff transformation
also generates a potential scattering term in the effective Kondo
model \cite{hewson}.

In the absence of an external magnetic field all above models preserve SU(2)
spin symmetry.
Spin anisotropies turn out to be irrelevant at the
critical fixed points, see Appendix~\ref{app:aniso}.
The effect of a magnetic field will be briefly discussed
in Sec.~\ref{sec:field}.


\subsection{Summary of results}

\begin{figure}
\epsfxsize=3.3in
\centerline{\epsffile{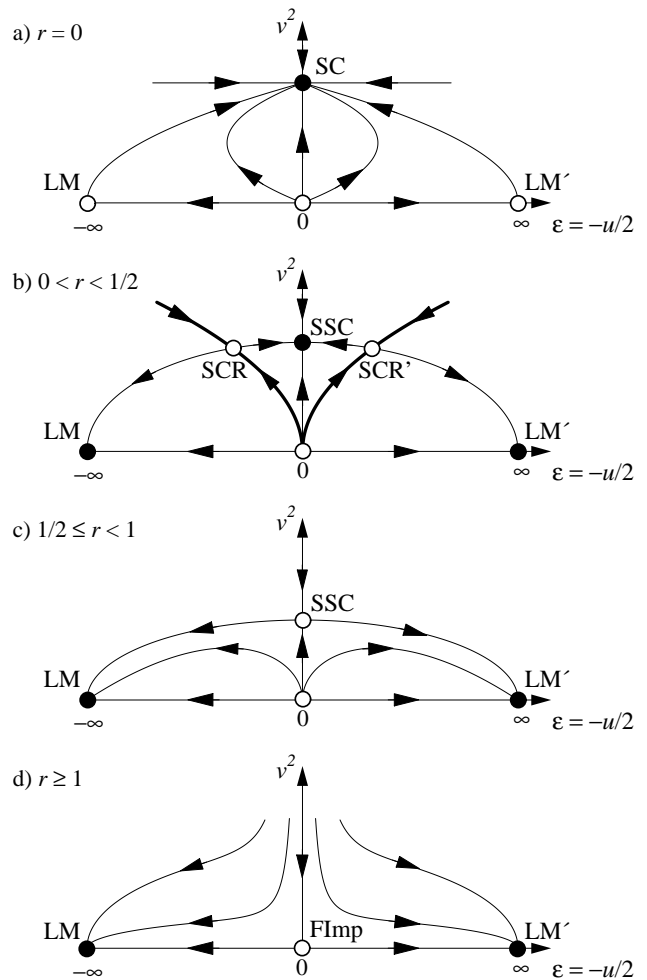}}
\caption{
Schematic RG flow diagrams for the particle-hole symmetric single-impurity
Anderson model with a pseudogap DOS, $\rho(\omega)\propto|\omega|^r$.
The horizontal axis denotes the renormalized on-site level energy $\epsf$
(related to the on-site repulsion $\U$ by $\U = -2 \epsf$),
the vertical axis is the renormalized hybridization $\hyb$.
The thick lines correspond to continuous boundary phase transitions;
the full (open) circles are stable (unstable) fixed points,
for details see text.
All fixed points at non-zero $\epsf$ have a mirror image at $-\epsf$, related
by the particle-hole transformation (\protect\ref{ph2}).
a) $r\!=\!0$, i.e., the familiar metallic case. For any finite $\hyb$
the flow is towards the strong-coupling fixed point (SC), describing Kondo screening.
b) $0\!<\!r\!<\!\frac{1}{2}$: The local-moment fixed point (LM) is stable,
and the transition to symmetric strong coupling (SSC) is controlled by the
SCR fixed point.
For $r\to 0$, SCR approaches LM, and the critical behavior at SCR is
accessible via an expansion in the Kondo coupling $j$.
In contrast, for $r\to\frac{1}{2}$, SCR approaches SSC, and the critical behavior can be
accessed by expanding in the deviation from SCR, i.e., in $\epsf=-\U/2$.
c) $\frac{1}{2}\!\leq\!r\!<\!1$: $\hyb$ is still relevant at $\U=0$. However, SSC is now
unstable w.r.t. finite $\U$.
At finite $\hyb$, the transition between the two stable fixed points
LM and LM' is controlled by SSC (which is now a critical fixed point!).
d) $r\!\geq\!1$: $\hyb$ is irrelevant, and the only transition is a level crossing
(with perturbative corrections) occurring at $\hyb=\U=0$, i.e., at the free-impurity
fixed point (FImp).
}
\label{fig:flowsym}
\end{figure}

\begin{figure}
\epsfxsize=3.2in
\centerline{\epsffile{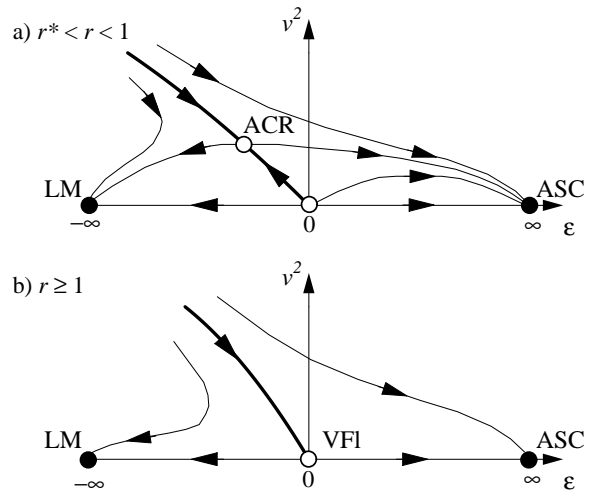}}
\caption{
Schematic RG flow diagram for the maximally particle-hole asymmetric
pseudogap Anderson impurity model.
The horizontal axis denotes the on-site impurity energy, $\epsf$,
the vertical axis is the fermionic coupling $\hyb$,
the bare on-site repulsion is fixed at $\U_0=\infty$.
The symbols are as in Fig.~\protect\ref{fig:flowsym}.
a) $r^\ast<r\!<\!1$: $\hyb$ is relevant, and the transition is controlled by an interacting
fixed point (ACR). 
As $r\to r^\ast \approx 0.375$,
p-h symmetry at the critical fixed point is dynamically restored,
and ACR merges into the SCR fixed point of Fig.~\protect\ref{fig:flowsym} -- this
cannot be described using the RG of Sec.~\protect\ref{sec:infu}.
In the metallic $r=0$ situation, studied by Haldane \protect\cite{haldane},
the flow from any point with $\hyb\neq 0$ is towards the screened singlet fixed
point with $\epsf=\infty$.
b) $r\!\geq\!1$: $\hyb$ is irrelevant, and the transition is a level crossing
with perturbative corrections, occuring at $\hyb=\epsf=0$, i.e., the
valence-fluctuation fixed point (VFl).
}
\label{fig:flowinfu}
\end{figure}

Our main results are summarized in the RG flow diagrams in
Figs.~\ref{fig:flowsym} and \ref{fig:flowinfu}, for the p-h symmetric and
asymmetric cases, respectively.

In the symmetric case, the ranges of exponent values $r=0$, $0<r<\frac{1}{2}$,
$\frac{1}{2}\leq r<1$, and $r\geq 1$ lead to quite different behavior,
and are shown separately in Fig.~\ref{fig:flowsym}.
No transition occurs for $r=0$: for any non-zero hybridization the flow is
towards the metallic Kondo-screened fixed point (SC).
This well-known fixed point can be identified as the stable fixed point of
a resonant level model; we argue below that this is actually an
intermediate-coupling fixed point.

For $0<r<\frac{1}{2}$, small values of the hybridization leave
the impurity spin unscreened provided that
$\epsfb< 0$, i.e., there is a stable local-moment fixed point (LM)
corresponding to $\epsf=-\infty$, $\hyb=0$.
A transition line at negative $\epsf$, with an unstable fixed point (symmetric critical, SCR)
at finite $\hyb$, $|\epsf|$, separates the flow towards LM from the flow to the
symmetric strong-coupling fixed point (SSC).
The strong-coupling fixed point displays its intermediate-coupling properties now
in a finite residual entropy and a finite magnetic moment, see
Sec.~\ref{sec:rlv}.
As $r\to 0$ the SCR fixed point merges with LM, in a manner characteristic
for a lower-critical dimension, i.e., with diverging correlation length
exponent.
A second critical fixed point SCR' exists for $\epsf>0$ which separates the
symmetric strong-coupling phase (SSC) from one with a free charge
doublet (LM').

As $r\to\frac{1}{2}$ the symmetric critical fixed points merge with the
strong-coupling one, again in a manner characteristic for a lower-critical
dimension.
For $r\geq\frac{1}{2}$ the fixed points SCR and SCR' cease to exist;
the strong-coupling SSC fixed point becomes infrared unstable,
and controls the LM -- LM' transition.
Finally, the structure of the flow changes again at $r=1$:
for $r\to 1$ the unstable strong-coupling fixed point (SSC) moves towards $\hyb=0$,
i.e., the free-impurity fixed point (FImp),
and for $r\geq 1$ no non-trivial fixed point remains.

For maximal p-h asymmetry, realized in the Anderson model through $\Ub=\infty$,
one has to distinguish exponent ranges $r=0$, $0<r\leq r^\ast$, $r^\ast<r<1$, and
$r\geq 1$.
In the metallic case $r=0$ any non-zero hybridization generates flow to strong
coupling with complete screening -- the strong-coupling fixed point is
the same as in the p-h symmetric situation, as p-h symmetry is marginally
irrelevant at strong coupling.
For all $r>0$ the situation is drastically different:
small $\hybb$ leaves the moment unscreened, whereas large
$\hybb$ directs the flow towards a new, p-h asymmetric, strong-coupling
fixed point (ASC).
The character of the critical fixed point separating the two phases
depends \cite{GBI} on $r$: for $0<r<r^\ast$ p-h symmetry is restored,
and the critical fixed point is the one of the p-h symmetric model.
For $r^\ast<r<1$ there is a separate critical fixed point (ACR)
which is p-h asymmetric,
i.e., located at finite $\hyb$ and $\epsf$.
For $r\to1$ the critical fixed point moves towards $\hyb\to 0$, and
for $r\geq 1$ the phase transition becomes a level crossing
(with perturbative corrections),
controlled by the valence-fluctuation fixed point (VFl),
see Fig.~\ref{fig:flowinfu}.

We finally discuss the general case of finite p-h asymmetry,
more details will be given in Sec.~\ref{sec:asym}.
Power counting shows that LM (SSC) are always (un)stable w.r.t. p-h asymmetry.
The symmetric critical SCR fixed point is stable w.r.t. p-h asymmetry for small $r$.
In contrast, for $r\lesssim\frac 1 2$ SCR is unstable towards p-h asymmetry,
as it is close to SSC in this regime.
This {\em requires} the existence of a specific $r$ value where this
change in character occurs: the is precisely $r=r^\ast\approx 0.375$ where p-h
asymmetry at SCR is marginal \cite{GBI}.
Upon increasing $r$ beyond $r^\ast$ the p-h asymmetric critical fixed point (ACR)
splits off from SCR.
In other words, upon approaching $r^\ast$ from large $r$ the ACR fixed point moves
towards small effective p-h asymmetry, and at $r=r^\ast$ ACR merges into SCR, implying
p-h symmetry is dynamically restored.
As stated above, the description of ACR using an expansion around VFl consequently
breaks down as $r\to {r^\ast}^+$.
Neither from numerics \cite{GBI} nor from the present RG are there indications
for the existence of a second asymmetric critical fixed point besides ACR;
thus, the critical properties for finite p-h asymmetry are always
equivalent to the ones of a model with maximal p-h asymmetry.

Taken together, the above observations show that $r=0$ plays the role of a lower-critical
dimension: as $r\to 0^+$, the correlation length exponent diverges, and
the second-order transition turns into a Kosterlitz-Thouless transition at $r=0$.
Interestingly, in the symmetric case the correlation length exponent also
diverges as $r\to \frac{1}{2}^-$, and the transition between LM and SSC
disappears for $r\geq\frac{1}{2}$: $r = \frac 1 2$ is a second lower-critical
dimension for the p-h symmetric problem.
In the asymmetric case, there is a transition between LM and ASC for all $r>0$,
and $r=1$ is equivalent to the upper-critical dimension, above which
the critical fixed point is non-interacting (actually a level crossing).


\subsection{Outline}

The rest of this paper is organized as follows:
Sec.~\ref{sec:obs} introduces the observables to be evaluated
in the course of the paper, together with their expected scaling
behavior near criticality.
In Sec.~\ref{sec:weak} we briefly review the standard weak-coupling
perturbative RG for the Kondo model, which is suitable to describe
the quantum phase transition for small $r$.
Sec.~\ref{sec:sym} discusses the particle-hole symmetric Anderson
model.
Starting from the non-interacting case, $\epsfb=\Ub=0$, we
first discuss the physics of the resulting non-interacting
resonant level model -- interestingly this can be identified
with a stable intermediate-coupling fixed point.
We then use a perturbative expansion in $\Ub$ to access the
critical fixed points for $r\lesssim\frac{1}{2}$.
In Sec.~\ref{sec:infu} we turn to the situation with maximal
p-h asymmetry, i.e., $\Ub=\infty$, and show that an expansion
in the hybridization provides access to the critical properties
for $r>1$ as well as for $r\lesssim 1$.
In Sec.~\ref{sec:asym} we consider the case of general
p-h asymmetry.
Sec.~\ref{sec:field} briefly describes the effect of a magnetic
field: the pseudogap model is shown to permit a sharp transition
as function of a field applied to the impurity for couplings
larger than the zero-field critical coupling.
In Sec.~\ref{sec:AvsK} we compare the physics of the Anderson
and Kondo models, arguing that the transitions in both models
fall in the same universality classes.
A brief discussion of applications concludes the paper.
All renormalization group calculations will employ the
field-theoretic RG scheme \cite{bgz} together with
dimensional regularization and minimal subtraction of poles,
with details given in the appendices;
one-loop RG results can equivalently be obtained using the
familiar momentum-shell method.


\section{Observables and scaling}
\label{sec:obs}

To establish notations and to pave the way for the RG analysis below,
we introduce a few observables together with their expected scaling
properties.

\subsection{Susceptibilities}

Magnetic susceptibilities are obtained by coupling an external magnetic field
to the bulk electronic degrees of freedom in ${\cal H}_{\rm b}$,
\begin{eqnarray}
&&- H_{\text{u}\alpha}(x) (c^\dagger_\sigma \sigma_{\sigma\sigma'}^\alpha c_{\sigma'})(x)
\end{eqnarray}
and to the impurity part $\mathcal{H}_{\text{A}}$, $\mathcal{H}_{\text{K}}$,
\begin{eqnarray}
- H_{\text{imp},\alpha} (f^\dagger_\sigma \sigma_{\sigma\sigma'}^\alpha f_{\sigma'})
\mbox{~,~~}
- H_{\text{imp},\alpha} {\hat S}_{\alpha}
\label{par2}
\end{eqnarray}
for the Anderson (\ref{aim}) and Kondo (\ref{pgk}) models, respectively.
The bulk field $H_{\text{u}}$ varies
slowly as function of the space coordinate, and $H_{\text{imp}}$ is the magnetic
field at the location of the impurity.

With these definitions, a spatially uniform field applied
to the whole system corresponds to $H_{\rm u} = H_{\rm imp} = H$.
Response functions can be defined from second derivatives of the thermodynamic
potential, $\Omega = - T \ln Z$, in the standard way \cite{vbs}:
$\chi_{\rm{u},\rm{u}}$ measures the bulk response to a field applied
to the bulk, $\chi_{\rm{imp},\rm{imp}}$ is the impurity response to
a field applied to the impurity, and $\chi_{\rm{u},\rm{imp}}$
is the cross-response of the bulk to an impurity field.

The impurity contribution to the total susceptibility is defined as
\begin{eqnarray}
\chi_{\rm imp}(T)
= \chi_{\rm imp,imp} + 2 \chi_{\rm u,imp} + (\chi_{\rm u,u} - \chi_{\rm u,u}^{\rm bulk})
\,,
\end{eqnarray}
where $\chi_{\rm u,u}^{\rm bulk}$ is the susceptibility of the bulk system in
absence of the impurity.
For an unscreened impurity spin of size $S=\frac{1}{2}$ we expect
$\chi_{\rm imp}(T\to 0) = 1/(4T)$ in the low-temperature limit,
and this is precisely the result in the whole LM phase.
A fully screened moment will be characterized by $T\chi_{\rm imp} = 0$;
note that the SSC fixed point displays a {\em finite} value of $T\chi_{\rm imp}$
for $r>0$.
At criticality $\chi_{\rm imp}$ does {\em not} acquire an anomalous dimension \cite{ss}
(in contrast to $\chi_{\rm loc}$ below),
because it is a response function associated to the conserved
quantity $S_{\rm tot}$.
Thus we expect a Curie law
\begin{equation}
\lim_{T\to 0} \chi_{\rm imp}(T) = \frac{C_{\rm imp}}{T} \,,
\label{fract}
\end{equation}
where the prefactor $C_{\rm imp}$ is in general a non-trivial universal constant
different from the free-impurity value $S(S+1)/3$.
Apparently, Eq.~(\ref{fract}) can be interpreted as the Curie response of a
fractional effective spin \cite{science}.

The local impurity susceptibility is given by
\begin{equation}
\chi_{\rm loc}(T) = \chi_{\rm imp,imp}  \,,
\label{chiloc}
\end{equation}
which is equivalent to the zero-frequency impurity
spin autocorrelation function.
In the unscreened phase we have $\chi_{\rm loc}\propto 1/T$ as $T\to 0$;
we can consider this as arising from the overlap
of the local impurity moment with the total, freely fluctuating, moment of $S=1/2$,
and so write
\begin{equation}
\lim_{T \rightarrow 0} \chi_{\text{loc}}(T) = \frac{m_{\text{loc}}^2}{4T} \,.
\label{mimp}
\end{equation}
The quantity $m_{\rm loc}$ turns out to be a suitable order parameter \cite{GBI,insi}
for the phase transitions between an unscreened and a screened spin:
it vanishes continuously as $t \to 0^-$,
where $t$ is the dimensionless measure of the distance
to criticality; in the Kondo model $t = (\Jb-J_c)/J_c$, whereas
in the Anderson model $t = (\hybb-{\hybb}_c) / {\hybb}_c$.
Thus, $T\chi_{\rm loc}$ is {\em not} pinned to the value of 1/4
for $t<0$ (in contrast to $T\chi_{\rm imp}$).
Remarkably, $m_{\rm loc}=0$ at the SSC fixed point for $r<1$,
although $T\chi_{\rm imp}=r/8$ there.

The phase transitions occurring for $0<r<1$ are described by
interacting fixed points, and thus obey strong hyperscaling properties,
including $\omega/T$ scaling in dynamical quantities \cite{book}.
For instance, the local dynamic susceptibility will follow a
scaling form
\begin{equation}
\chi''_{\rm loc}(\omega,T)
= \frac {{\cal B}_1} {\omega^{1-\eta_{\chi}}} \, \Phi_1 \!\left(\frac{\omega}{T}, \frac{T^{1/\nu}}{t}\right)
\label{scalchi1}
\end{equation}
which describes critical local-moment fluctuations,
and the local static susceptibility obeys
\begin{equation}
\chi_{\rm loc}(T)
= \frac {{\cal B}_2} {T^{1-\eta_{\chi}}} \, \Phi_2 \!\left(\frac{T^{1/\nu}}{t}\right) \,.
\label{scalchi2}
\end{equation}
Here, $\eta_{\chi}$ is a universal anomalous exponent,
which controls the anomalous decay of the
two-point correlations of the impurity spin,
and $\Phi_{1,2}$ are universal crossover functions (for the specific critical
fixed point and for fixed $r$), whereas
${\cal B}_{1,2}$ are non-universal prefactors.
Furthermore, $\nu$ is the correlation length exponent, describing
the flow away from criticality:
when the system is tuned through the transition,
the characteristic energy scale $T^\ast$, above which critical behavior
is observed, vanishes as \cite{book}
\begin{equation}
T^\ast \propto |t|^{\nu} \,;
\end{equation}
the dynamical critical exponent $z$ can be set to unity in the
present $(0+1)$-dimensional problem.
Note that at criticality, $t=0$, the relation (\ref{scalchi2}) reduces to
$\chi_{\rm loc}(T) \propto T^{-1+\eta_\chi}$.

Hyperscaling can be used to derive relations between critical exponents.
The susceptibility exponent $\eta_\chi$ and the correlation length exponent $\nu$
of a specific transition
are sufficient to determine all critical exponents associated with
a local magnetic field \cite{insi}.
In particular, the $T\to 0$ local susceptibility away from criticality
obeys
\begin{eqnarray}
\chi_{\rm loc}(t>0)  &\propto& t^{-\gamma} \,,~
\gamma = \nu \,(1-\eta_\chi) \,,\nonumber\\
T\chi_{\rm loc}(t<0) &\propto& (-t)^{\gamma'} \,,~
\gamma' = \nu \, \eta_\chi \,,
\end{eqnarray}
which can be derived from a scaling ansatz for the impurity part of the
free energy \cite{insi}.
The last relation implies the order parameter vanishing as
\begin{equation}
m_{\rm loc} \propto (-t)^{\nu\eta_\chi /2} \,.
\end{equation}
Note that hyperscaling holds for all critical fixed points of the
pseudogap Kondo problem with $0<r<1$.

\subsection{Impurity entropy}

In general, zero-temperature critical points in quantum impurity models
can show a finite residual entropy [in contrast
to bulk quantum critical points where the entropy usually vanishes
with a power law, $S(T) \propto T^y$].
For the models at hand, the impurity contribution to the
low-temperature entropy is obtained by a perturbative evaluation
of the thermodynamic potential and taking the temperature
derivative.
This will yield epsilon-type expansions for the
ground state entropy $S_{\rm imp}(T=0)$, with explicit results
given below.

Note that the impurity part of the thermodynamic potential will usually
diverge with the cutoff, i.e., we have
$\Omega_{\rm imp} = E_{\rm imp} - T S_{\rm imp}$,
where $E_{\rm imp}$ is the non-universal (cutoff-dependent) impurity contribution
to the ground-state energy.
However, the impurity entropy $S_{\rm imp}$ is fully universal, and
the UV cutoff can be sent to infinity {\em after} taking the
temperature derivative of $\Omega_{\rm imp}$.

Thermodynamic stability requires that the total entropy of a system decreases
upon decreasing temperature, $\partial_T S(T) > 0$.
This raises the question of whether the impurity part of the entropy, $S_{\rm imp}$
has to decrease under RG flow (which is equivalent to decreasing $T$).
The so-called $g$-theorem \cite{gtheorem} provides a proof of this conjecture
for systems with short-ranged interactions;
for most quantum impurity problems this appears to apply.
Interestingly, the pseudogap Kondo problem provides an explicit counter-example,
as the two critical fixed points obey $S_{\rm SCR} < S_{\rm ACR}$, with the RG
flow being from SCR to ACR (!), see Sec.~\ref{sec:infu} for details.
(For another counter-example see Ref.~\onlinecite{uphill}.)

\subsection{$T$ matrix}

An important quantity in an Anderson or Kondo model
is the conduction electron $T$ matrix, describing the scattering of
the $c$ electrons off the impurity.
For an Anderson model, the $T$ matrix is just given by
$T(\omega) = \hybb^2 G_f(\omega)$ where $G_f$ is the full impurity
$f$ electron Green's function.
For a Kondo model, it is useful to define a propagator $G_T$ of the
composite operator $T_\sigma = f_\sigma^\dagger f_{\sigma'} c_{\sigma'}$,
such that the $T$ matrix is given by $T(\omega) = \Jb^2 G_T(\omega)$.
As with the local susceptibility, we expect a scaling form of the
$T$ matrix spectral density near the intermediate-coupling fixed points
similar to Eq.~(\ref{scalchi1}).
In particular, at criticality a power law occurs:
\begin{equation}
T(\omega) \propto \frac{1}{\omega^{1-\eta_T}} \,.
\end{equation}
Remarkably, we will find the {\em exact} result $\eta_T = 1-r$ for $r<1$, i.e.,
for all {\em interacting} critical fixed points considered in this paper
the $T$ matrix follows $T(\omega) \propto \omega^{-r}$.
NRG calculations have found precisely this critical divergence,
for both the symmetric and asymmetric critical points \cite{bulla,MVRB}.
At the trivial fixed points (LM, ASC) the behavior of the $T$ matrix
follows from perturbation theory in the hybridization to be
$T(\omega)\propto\omega^r$.

Notably, the $T$ matrix can be directly observed in experiments,
due to recent advances in low-temperature scanning tunneling microscopy,
as has been demonstrated, e.g., with high-temperature
superconductors \cite{seamus,tolya,MVRB}.

\subsection{Phase shifts}

Fixed points which can be described in terms of free fermions
can be characterized by the $s$-wave conduction electron phase shift, $\delta_0(\omega)$,
which can be related to the conduction electron $T$ matrix
through $\delta_0(\omega) = {\rm arg} T(\omega)$.
A decoupled impurity simply has a phase shift $\delta_0 = 0$,
whereas a p-h symmetric Kondo-screened impurity in a metallic host has
a low-energy phase shift of $\delta_0(\omega) = \frac\pi 2 {\rm sgn}(-\omega)$.
A detailed discussion for the pseudogap model has been given in Ref. \onlinecite{GBI},
in the body of the paper we will simply quote the results.


\section{Weak-coupling RG for the Kondo model}
\label{sec:weak}

In this section we briefly summarize the weak-coupling
RG for the pseudogap Kondo model (\ref{pgk}), as first
discussed by Withoff and Fradkin \cite{withoff}.
Perturbative RG is performed around $\Jb=0$, i.e., the
local-moment fixed point (LM):
this will allow to access the (p-h symmetric) critical fixed
point SCR, which is located close to LM for small DOS exponents $r$.

\subsection{Lower critical dimension: Expansion around the local-moment fixed point}
\label{sec:rgweak}

The RG flow equation for the renormalized Kondo coupling $j$,
to two-loop order, reads \cite{MKMV}
\begin{equation}
\label{betaj}
\beta(j) = r j - j^2 + \frac{j^3}{2} \,.
\end{equation}
This yields an infrared unstable fixed point at
\begin{equation}
\label{jfix}
j^\ast = r + \frac{r^2}{2} + {\cal O}(r^3)
\end{equation}
which controls the transition between the decoupled LM and the
Kondo-screened SSC phases.
The small-$j$ expansion (\ref{betaj}) -- which is nothing but
the generalization of Anderson poor man's scaling \cite{poor}
to the pseudogap case --
cannot give information about
the strong-coupling behavior, and it can only describe critical
properties for small $r$.
(In the p-h symmetric case, the fixed point structure does not change
within the exponent range $0<r<\frac 1 2$, thus the present expansion is
in principle valid up to $r=\frac 1 2$.)

Adding a potential scattering term $V_0$ gives a finite p-h asymmetry.
Under RG, we find that $V$ renormalizes to zero for $r>0$,
$\beta(V) = rV$.
Thus, within the range of applicability of the weak-coupling RG,
p-h asymmetry is irrelevant.
(Strictly, this applies for $r<r^\ast$, see Ref.~\onlinecite{GBI}.)

\subsection{Observables near criticality}
\label{sec:weakobs}

We quote a few properties of the critical regime which
have been determined in Ref.~\onlinecite{MKMV}.
Expanding the beta function (\ref{betaj}) around the fixed point value (\ref{jfix})
yields the correlation length exponent $\nu$:
\begin{equation}
\frac{1}{\nu} = r - \frac{r^2}{2} +  {\cal O}(r^3)\,.
\label{nuz_weak}
\end{equation}
The low-temperature impurity susceptibility and entropy at criticality
are given by
\begin{eqnarray}
\label{tchi_weak}
T\chi_{\rm imp} &=& \frac{1}{4} (1 - r) + {\cal O}(r^2)\,, \\
\label{simp_weak}
S_{\rm imp} &=& \ln 2 \bigg(1 + \frac{3\pi^2}{8}\,r^3\bigg) + {\cal O}(r^5) \,.
\end{eqnarray}
The anomalous exponent of the local susceptibility evaluates to
\begin{equation}
\label{etachi_weak}
\eta_\chi = r^2 + {\cal O}(r^3) \,.
\end{equation}
A comparison of the above results with numerical data is given in Figs.~\ref{fig:nuz},
\ref{fig:etachi}, \ref{fig:tchi}, and \ref{fig:simp} below.

Most importantly, the continuous transition controlled by the fixed point (\ref{jfix}),
which exists only for $r>0$, evolves smoothly into the
Kosterlitz-Thouless transition at $r=0$, $j=0$, which separates the
antiferromagnetic and the ferromagnetic metallic Kondo model.
This is also indicated by the divergence of the correlation length
exponent (\ref{nuz_weak}) as $r\to 0^+$.
Thus, $r=0$ can be identified as a lower-critical ``dimension'' of
the pseudogap Kondo problem.
It is interesting to compare the present expansion with the $(2+\epsilon)$ expansion
for the non-linear sigma model, appropriate for magnets close to the
lower-critical dimension.
The expansion is done about the ordered magnet,
thus the LM phase with ln 2 residual entropy takes the role of the
{\em ordered} state in the pseudogap Kondo problem.


\section{Particle-hole symmetric Anderson model}
\label{sec:sym}

In the following sections of the paper
we shift our attention from the Kondo model
to the impurity Anderson model with a pseudogap density of states.
This formulation will provide new insights into
the RG flow and the critical behavior of both the Anderson and Kondo
models.

The coupling between impurity and bath is now the Anderson
hybridization term, $\hybb$, which turns out to be marginal in
a RG expansion around $\hybb=0$ for the DOS exponent $r=1$
(in contrast to the Kondo coupling $\Jb$ which is marginal for $r=0$).
As we will show in Sec.~\ref{sec:infu},
the Anderson model provides the relevant
low-energy degrees of freedom for the p-h asymmetric pseudogap
transition near its upper-critical dimension.

Interestingly, also the p-h symmetric version of the Anderson model
allows to uncover highly non-trivial physics, in particular
the special role played by the DOS exponent $r=\frac{1}{2}$,
where the transition disappears in the presence of p-h symmetry.
Thus we start our analysis with the particle-hole symmetric
Anderson model (\ref{aim}),
i.e., we keep $\Ub=-2\epsfb$ and discuss the physics as
function of $\hybb$ and $\epsfb$.

\subsection{Trivial fixed points}
\label{sec:symtriv}

For vanishing hybridization, $\hybb=0$, the symmetric Anderson model (\ref{aim})
features three trivial fixed points:
for $\epsfb<0$ the ground state is a spinful
doublet -- this represents the local-moment fixed point (LM).
For $\epsfb>0$ we find a doublet of states (empty and doubly occupied),
denoted as LM' and related to LM by the p-h transformation (\ref{ph2}).
Both LM and LM' have a residual entropy of $S_{\rm imp} = \ln 2$.
At $\epsfb=0$ a level crossing between the two doublets occurs, i.e.,
all four impurity states are degenerate --
this is the free-impurity fixed point (FImp), with residual entropy $\ln 4$.
The impurity spin susceptibilities are
\begin{equation}
T\chi_{\rm imp} =
\left\{
\begin{array}{ll}
1/4 & \mbox{LM} \\
1/8 & \mbox{FImp} \\
 0  & \mbox{LM'} \\
\end{array}
\right. ,
\end{equation}
the conduction electron phase shift is zero at all these fixed points.
The hybridization term, $\hybb$, is irrelevant at both the LM and LM' fixed points
for $r>0$, whereas for $r=0$ it is marginally relevant, as shown by
the RG in Sec.~\ref{sec:rgweak}.

\subsection{Resonant level model: Intermediate-coupling fixed point}
\label{sec:rlv}

It proves useful to discuss the $\epsfb=\Ub=0$ case,
i.e., the physics on the vertical axis of the flow diagrams
in Fig.~\ref{fig:flowsym}.
This non-interacting system is known as resonant level model, as
the two spin species are decoupled.
The problem can be solved exactly: the $f$ electron self-energy
is
\begin{equation}
\Sigma_f = \hybb^2 G_{c0}
\label{sig_rl}
\end{equation}
where $G_{c0}$ is the bare conduction electron Green's function
at the impurity location ${\bf R}=0$.
In the low-energy limit the $f$ electron propagator is then given by
\begin{equation}
G_f(i\omega_n)^{-1} = i\omega_n - i A_0\,{\rm sgn}(\omega_n)\,|\omega_n|^r
\label{dressedf}
\end{equation}
where the $|\omega_n|^r$ self-energy term dominates for $r<1$,
and the prefactor $A_0$ is
\begin{equation}
\label{A0}
A_0 = \frac{\pi V_0^2}{\cos\frac{\pi r}{2}} \,.
\end{equation}

Before stating results for observables it is interesting to tackle
the problem using RG techniques,
with an expansion in the hybridization strength $\hybb$
around the free-impurity fixed point (FImp, $\hybb=0$).
We study the action
\begin{eqnarray}
{\cal S} = \int_0^\beta d \tau
&\bigg[&
\left[
\bar{f}_\sigma \partial_\tau f_\sigma + \hybb (\bar{f_\sigma}c_{\sigma}(0) + {\rm c.c.})
\right] \nonumber \\
&+&
\int_{-\Lambda}^{\Lambda} d k |k|^r \,
\bar{c}_{k\sigma} (\partial_\tau - k)
c_{k\sigma}
\bigg]
\label{rlvact}
\end{eqnarray}
where $c_{\sigma}(0)$ is the bath fermion field at the impurity position
as above.
Power counting w.r.t. the $\hybb=0$ fixed point, using
${\rm dim}[f] = 0$, ${\rm dim}[c(0)] = (1+r)/2$,
yields
\begin{equation}
\label{dimhyb}
{\rm dim}[\hybb] = \frac{1-r}{2} \equiv \bar{r} \,,
\end{equation}
i.e., the hybridization is relevant only for $r<1$.

To perform RG within the field-theoretic scheme \cite{bgz},
we introduce a renormalized hybridisation $\hyb$ according
to $\hybb = (Z_\hyb \mu^{\bar{r}}/\sqrt{Z}) \hyb$,
where $\mu$ is a renormalization energy scale, and
$Z_\hyb$ and $Z$ are the interaction and field renormalization factors.
The RG flow equation for $\hyb$ is found to be
\begin{equation}
\label{betagsym}
\beta(\hyb) = - {\bar r} \hyb +  \hyb^3 \,.
\end{equation}
Remarkably, this result is {\em exact} to all orders in perturbation
theory: the cubic term arises from the {\em only} self-energy diagram of
the $f$ fermions, and no vertex renormalizations occur ($Z_\hyb=1$).
This implies that the low-energy physics of the non-interacting resonant level
model is controlled by the stable intermediate-coupling fixed point
located at
\begin{equation}
{\hyb^\ast}^2 = \bar{r} = \frac{1-r}{2}
\label{g_rlv}
\end{equation}
for $0\leq r<1$, which also applies to the familiar metallic case $r=0$.

The intermediate-coupling nature of the stable fixed point, with
associated universal properties, is consistent with the results
known from the exact solution of the problem, e.g., a universal
conduction electron phase shift, a universal crossover in the
temperature-dependent susceptibility etc.

We proceed with calculating a number of observables for the pseudogap
resonant level model.
Interestingly, this can be done in two ways:
either (i) via the exact solution of the problem, i.e., by
integrating out the $c$ fermions exactly [leading to the propagator (\ref{dressedf})],
or equivalently (ii) by evaluating perturbative corrections to the FImp
fixed point using the RG result (\ref{g_rlv}), utilizing standard
renormalized perturbation theory, and noting that all corrections beyond
second order in $\hyb$ vanish exactly within this scheme.
Details of the calculation are in Appendix~\ref{app:rl}.

We start with evaluating spin susceptibilities -- note that we have kept two spin
species in the model.
The zero-temperature dynamic local susceptibility is proportional to the
bubble formed with two $f$ propagators (\ref{dressedf}),
\begin{equation}
\chi''_{\rm loc}(\omega) \propto
\left\{
\begin{array}{ll}
\omega^{1-2r}                    &  (0 \leq r<1) \\
\delta(\omega)\frac{\omega}{T}   &  (r \geq 1) \\
\end{array}
\right. \,,
\end{equation}
where the case of $r=\frac{1}{2}$ receives logarithmic corrections, see below.
The low-temperature limit of the impurity susceptibility is found to be
\begin{equation}
\label{tchi_ssc}
T\chi_{\rm imp}(T) = \frac{r}{8} \,,
\end{equation}
the impurity entropy is
\begin{equation}
\label{simp_ssc}
S_{\rm imp} = 2 r \ln 2 \,,
\end{equation}
where the two last equations are valid for $0\leq r < 1$;
for $r\geq 1$ the resonant level model flows to the free-impurity fixed point (FImp)
with properties listed in Sec.~\ref{sec:symtriv}.
The conduction electron phase shift near the Fermi level,
determined in Ref.~\onlinecite{GBI}, is
\begin{equation}
\frac{\delta_0(\omega)}{{\rm sgn}(-\omega)} =
\left\{
\begin{array}{ll}
(1-r)\frac\pi 2                    &  (0 \leq r<1) \\[2mm]
\frac{\pi}{2\ln|\Lambda/\omega|}   &  (r=1) \\[2mm]
{\cal O}(\omega^{r-1})             &  (r>1)
\end{array}
\right. \,.
\end{equation}

Interestingly, the resonant level model describes a screened impurity
only in the metallic case.
For the pseudogap case, $r>0$, Eqs.~(\ref{tchi_ssc}) and (\ref{simp_ssc})
show that the impurity is only partially screened:
in a model of free fermions we have a residual entropy!
We will see below that the resonant level
model fixed point (\ref{g_rlv}) can be identified with the symmetric
strong-coupling fixed point (SSC) of Gonzalez-Buxton and Ingersent \cite{GBI} ,
introduced for the p-h symmetric Kondo and Anderson models.

\subsection{Expansion around the resonant level fixed point}
\label{sec:SSCRG}

After having analyzed the behavior of the Anderson model
in the non-interacting case, we proceed to study the stability
of the resonant level fixed point w.r.t. a finite interaction
strength $\Ub$, keeping p-h symmetry.
Importantly, this fixed point, characterized by a finite hybridization strength
between impurity and bath, is stable for small $r$ [see Eq.~(\ref{dimu})];
we conclude that it can be identified with the SSC fixed point of Ref.~\onlinecite{GBI}.

Numerical results \cite{bulla,GBI} indicate that the quantum phase transition
between LM and SSC disappears as $r$ is increased to $\frac{1}{2}$,
where the p-h symmetric critical fixed point (SCR) merges with the SSC fixed
point.
We shall show that an expansion around the SSC fixed point captures the
physics of the SCR fixed point for $r\lesssim \frac{1}{2}$.
Thus, this expansion describes {\em the same} critical fixed point as the
weak-coupling expansion of Sec.~\ref{sec:weak}, but approaching it from $r=\frac{1}{2}$
instead of $r=0$.
(The $r$ values 0 and $\frac 1 2$ are {\em two} lower-critical dimensions for the
p-h symmetric pseudogap Kondo problem.)

The RG expansion below will be performed around an {\em intermediate-coupling}
fixed point, in contrast to most analytical RG calculations which expand
around trivial (i.e. weak or strong-coupling) fixed points.
Strategically,
one could think about a double expansion in $\hybb$ and $\Ub$.
However, this is not feasible, as the marginal dimensions for both
couplings are different, $r=1$ and $r=\frac{1}{2}$, respectively.
Therefore we choose to first integrate out the $c$ fermions exactly,
and then use standard RG tools for the expansion in $\Ub$.

Consequently, the starting point is the action
\begin{eqnarray}
\label{symact}
{\cal S} &=&
  \sum_{\omega_n} \bar{f}_\sigma(\omega_n)\,[i A_0 {\rm sgn}(\omega_n)|\omega_n|^r] \,f_\sigma(\omega_n) \\
&+&
  \int_0^\beta d \tau \,
  U_0 \bigg(\bar{f}_\uparrow f_\uparrow-\frac{1}{2}\bigg) \bigg(\bar{f}_\downarrow f_\downarrow-\frac{1}{2}\bigg)
\nonumber
\end{eqnarray}
where the $f$ fermions are now ``dressed'' by the conduction lines.
$A_0$ is the non-universal number given in Eq. (\ref{A0}),
and we have assumed $0<r<1$.
The interaction term has been written in a p-h symmetric form.
A note is in order regarding the cutoff:
The original model had a UV cutoff $\Lambda$, and this sets the
upper bound for the $|\omega_n|^r$ behavior of the $f$ propagator (\ref{dressedf}),
i.e., $\Lambda$ is now the energy cutoff for the spectral density of the $f$
fermions.
The RG to be performed below can be understood as progressive reduction
of this cutoff
(although we will use the field-theoretic scheme where the cutoff is
implicitely sent to infinity at an early stage).

Dimensional analysis w.r.t. the $\Ub=0$ situation, using
${\rm dim}[f] = (1-r)/2$, results in
\begin{equation}
\label{dimu}
{\rm dim}[U_0] = 2r-1 \equiv -\epsilon\,,
\end{equation}
hence, for $r>1/2$ the interaction term is relevant and
the SSC fixed point is unstable.

We continue with the RG analysis of (\ref{symact}).
To perform a perturbative expansion in $\Ub$ using the field-theoretic
scheme \cite{bgz}, we introduce a renormalized field and a
dimensionless coupling according to
\begin{eqnarray}
f_\sigma &=& \sqrt{Z} \, f_{R\sigma} \,,  \\
\Ub &=& \frac{\mu^{-\epsilon} A_0^2 Z_4} {Z^2} \U \label{udef}
\end{eqnarray}
where $\mu$ is a renormalization energy scale as usual, and
$(-\epsilon)$
is the bare scaling dimension of $\Ub$;
we have absorbed the non-universal number $A_0$ appearing in the dynamic term
of the action (\ref{symact}) in order to obtain a universal fixed point value of $u$.

The complete RG analysis, needed to determine the flow of $\U$, is presented
in Appendix~\ref{app:rgsym}, here we restrict ourselves to the final
results.
P-h symmetry prohibits the occurence of even powers
of $\U$ in the beta function of $\U$,
therefore the lowest contributions arise at two-loop order.
Remarkably, no singular propagator renormalizations occur,
thus
\begin{equation}
Z = 1
\label{zeq1}
\end{equation}
to all orders in perturbation theory.
The RG flow equation for the renormalized interaction $\U$,
arising now only from two-loop vertex renormalizations,
is found to be
\begin{equation}
\label{betau}
\beta(\U) = \epsilon \U -  \frac{3(\pi-2 \ln 4)}{\pi^2} \U^3 \,.
\end{equation}
Next-to-leading order contributions would require a four-loop
calculation which we do not attempt here.
For positive $\epsilon$, i.e., $r<\frac 1 2$,
Eq.~(\ref{betau}) yields a pair of unstable fixed points at
finite $|\U|$ (in addition to the stable one at $\U=0$);
the correlation length exponent of the transition, Eq.~(\ref{nuz_sym}),
diverges as $r\to\frac{1}{2}^-$.
Thus, the behavior {\em below} $r=\frac 1 2$ is similar to the
standard behavior {\em above} a lower-critical dimension
(e.g. in the non-linear sigma model for bulk magnet case).

\subsection{$r=0$}

Clearly, the unstable finite-$\U$ fixed points predicted by the perturbative
RG equation (\ref{betau}) do not necessarily exist for the metallic case
$r=0$, as $\epsilon = 1$ is possibly outside the convergence radius of the
expansion.
Indeed, the Kondo RG of Sec.~\ref{sec:rgweak} shows that, at $r=0$,
the LM and LM' fixed points are unstable w.r.t. finite impurity coupling.
As the resonant-level fixed point at $\U=0$ is stable,
we conclude that the flow is directly from
LM (LM') to SC, and SC represents the familiar strong-coupling Kondo
fixed point, with complete screening of the spin.
The RG flow is in Fig.~\ref{fig:flowsym}a.

\subsection{$0<r<\frac{1}{2}$}

For $r$ values smaller than $\frac{1}{2}$, both the $\hyb=\hyb^\ast$,$\U=0$ fixed point (SSC)
and the $\hyb=0$, $|\epsf|=\infty$ fixed points (LM, LM') are stable,
and should be separated by critical fixed points.
The RG equation (\ref{betau}) yields a pair of infrared unstable fixed points
at
\begin{equation}
{\U^\ast}^2 = \frac{\pi^2}{3(\pi-2\ln 4)} \epsilon + {\cal O}(\epsilon^2)
\label{u_scr}
\end{equation}
with $\epsilon=1-2r$.
These two fixed points represent SCR and SCR', see the flow diagram in
Fig.~\ref{fig:flowsym}b.
Note that p-h symmetry also dictates that the flow trajectories out of the
SSC fixed point are horizontal in the $\U$-$\hyb^2$ diagram.
Therefore, close to $r=\frac 1 2$ the SCR and SCR' fixed points are completely
described by the fixed point coupling values
$\hyb^\ast$ (\ref{g_rlv}) and $\U^\ast$ (\ref{u_scr}).

\subsection{$\frac{1}{2} \leq r < 1$}

For $r > \frac{1}{2}$ ($r=\frac{1}{2}$) the self-interaction $\U$
is a (marginally) relevant perturbation at the SSC fixed point,
and (\ref{betau}) does not yield additional non-trivial fixed
points.
As LM and LM' are stable, we can conclude that the flow
is from SSC directly towards LM (LM') for positive (negative) $\Ub$.
Hence, SSC has become a critical fixed point, controlling
the transition between LM and LM', which occurs at
$\Ub = 0$ for any finite $\hybb$.
We shall not consider this transition in greater detail,
apart from stating its correlation length exponent,
$1/\nu = -\epsilon$.
Fig.~\ref{fig:flowsym}c displays the flow diagram arising from this discussion,
being consistent with the numerical results of Ref.~\onlinecite{GBI}.

\begin{figure}
\epsfxsize=3.1in
\centerline{\epsffile{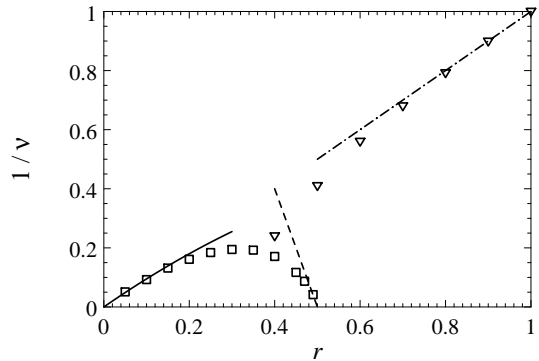}}
\caption{
Inverse correlation length exponent $1/\nu$ obtained from NRG,
at both the symmetric (squares) and asymmetric (triangles) critical points,
together with the analytical RG results from the expansions
in $r$ [Sec.~\protect\ref{sec:weak}, Eq.~(\protect\ref{nuz_weak}), solid],
in $(\frac{1}{2} -r)$ [Sec.~\protect\ref{sec:sym}, Eq.~(\protect\ref{nuz_sym}), dashed], and
in $(1-r)$ [Sec.~\protect\ref{sec:infu}, Eq.~(\protect\ref{nuz_infu}), dash-dot].
The numerical data have been partially extracted from Ref.~\protect\onlinecite{insi}
using hyperscaling relations; for the symmetric model data are from
Ref.~\protect\onlinecite{bulla}.
}
\label{fig:nuz}
\end{figure}

\subsection{$r\geq 1$}

The physics of the symmetric Anderson model for $r\geq 1$ is easily discussed:
the hybridization term $\hybb$ is irrelevant for all $\Ub$,
Eq.~(\ref{dimhyb}).
The free-impurity fixed point (FImp) is the only remaining fixed point
at $\Ub=0$.
It is unstable w.r.t. finite $\Ub$, and controls the transition between the
two stable fixed points LM and LM'.
The resulting flow diagram is in Fig.~\ref{fig:flowsym}d.

\subsection{Observables near criticality}
\label{sec:symobs}

Here we discuss critical properties of the SCR fixed point,
the properties of SCR' are identical when translated from spin to
charge degrees of freedom.
The correlation length exponent follows from expanding the beta function
(\ref{betau}) around its fixed-point value:
\begin{equation}
\label{nuz_sym}
\frac{1}{\nu} = 2-4r + {\cal O}(\epsilon^2)
\end{equation}
A comparison with results from NRG is shown in Fig.~\ref{fig:nuz}.
Close to $r=\frac 1 2$, the analytical expression nicely matches the numerical
results, however, higher-order corrections in the expansion quickly become
important.

We continue with the quantities introduced in Sec.~\ref{sec:obs}.
As usual for an expansion where the non-linear coupling
has an infrared {\em unstable} fixed point
(as occurs above the lower-critical dimension in standard situations),
the UV cutoff needs to be kept explicitly, and
intermediate quantities will diverge with the UV cutoff
(see e.g. the calculation of the impurity entropy).
However, these divergences will cancel in the final
expressions for universal observables, and this is
an important check for the consistency of our
calculations.

\begin{figure}[t]
\epsfxsize=3.3in
\centerline{\epsffile{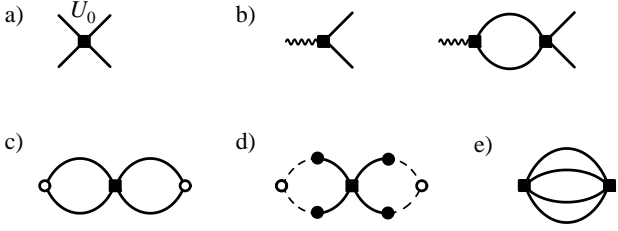}}
\caption{
Feynman diagrams for the symmetric Anderson impurity model,
where the expansion is done in $\Ub$ around the resonant level model
fixed point.
Full lines are {\em dressed} $f_\sigma$ propagators with the self-energy
arising from the conduction electrons already taken into account,
i.e., with the propagator given in Eq.~(\protect\ref{dressedf}).
a) Bare interaction vertex $\Ub$.
b) Correlation function involving a composite $(\bar{f}f)$ operator (wiggly line),
together with the first perturbative correction.
c) $\Ub$ correction to the local susceptibility $\chi_{\rm imp,imp} = \chi_{\rm loc}$,
open circles are sources.
d) $\Ub$ correction to $\chi_{\rm u,u}$, which contributes to the
Curie term of the impurity susceptibility $\chi_{\rm imp}$.
Dashed lines are conduction electron lines, and the full dots
are the $\hybb$ vertices.
e) $\Ub^2$ contribution to the impurity free energy.
}
\label{fig:symdgr1}
\end{figure}

\subsubsection{Local susceptibility}

The local susceptibility at the SSC fixed point, i.e., at tree level,
follows the power law
$\chi_{\rm loc} \propto \omega^{1-2r} = \omega^\epsilon$.
To obtain corrections to the tree-level result,
one introduces a $\chi_{\rm loc}$ renormalization factor, $Z_\chi$,
from which one obtains the anomalous exponent according to
\begin{equation}
\eta'_\chi = \beta(\U) \frac{d \ln Z_\chi}{d\U} \bigg|_{\U^\ast} \,.
\end{equation}
Note that $1-\eta_\chi = \epsilon + \eta'_\chi$, with
$\eta_\chi$ defined in Eqs. (\ref{scalchi1},\ref{scalchi2}),
due to the non-trivial structure of the problem already at tree level.

\begin{figure}
\epsfxsize=3.1in
\centerline{\epsffile{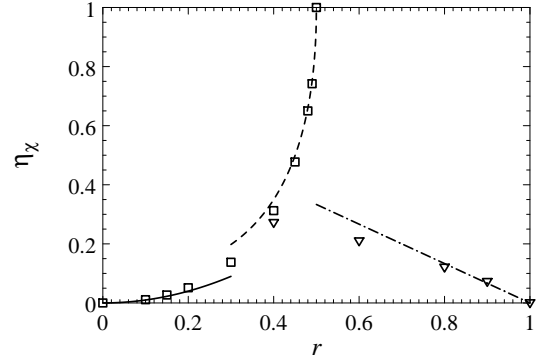}}
\caption{
NRG data \cite{insi} for the local susceptibility exponent, $\eta_\chi$, defined through
$\chi_{\rm loc} \propto T^{-1+\eta_\chi}$,
at both the symmetric (squares) and asymmetric (triangles) critical points,
together with the renormalized perturbation theory results from the expansions
in $r$ [Sec.~\protect\ref{sec:weak}, Eq.~(\protect\ref{etachi_weak}), solid],
in $(\frac{1}{2} -r)$ [Sec.~\protect\ref{sec:sym}, Eq.~(\protect\ref{etachi_sym}), dashed], and
in $(1-r)$ [Sec.~\protect\ref{sec:infu}, Eq.~(\protect\ref{etachi_infu}), dash-dot].
}
\label{fig:etachi}
\end{figure}

Different ways can be used to determine $Z_\chi$.
Realizing that $\chi_{\rm loc}$ is a correlation function of a composite
$(\bar{f}f)$ operator leads to $Z_\chi = Z_2^2$, where
$Z_2$ is the renormalization factor associated with $(\bar{f}f)$.
$Z_2$ can be calculated from the correlation function shown in
Fig.~\ref{fig:symdgr1}b, which receives a perturbative correction to first
order in $\Ub$.
Alternatively, $\chi_{\rm loc}$ can be calculated directly,
and a single diagram gives a contribution of order $\Ub$ (Fig.~\ref{fig:symdgr1}c).
Both ways lead to
\begin{equation}
\label{zchisym}
Z_\chi = 1 + \frac{2\U}{\pi\epsilon} \,,
\end{equation}
with details given in Appendix~\ref{app:rgsym}.
The result for $\chi_{\rm loc}$ is proportional to the non-universal
number $A_0^{-2}$, however, the exponent is universal:
\begin{equation}
\eta'_\chi = \frac{2}{\pi} \U^\ast = 2 [3(\pi-2\ln 4)]^{-1/2} \sqrt{\epsilon}
\end{equation}
The local susceptibility thus follows $\chi_{\rm loc}(T) \propto T^{-1+\eta_\chi}$ with
\begin{equation}
\label{etachi_sym}
\eta_\chi = 2-2r \,+\, 2.688 \sqrt{\frac{1}{2} -r} \,+\, {\cal O}(\epsilon)
\end{equation}
where the first term contains the tree level expression, and further ${\cal O}(\epsilon)$
terms arise from higher-order perturbative corrections.
A comparison with NRG results is given in Fig.~\ref{fig:etachi}, where good
agreement near $r=\frac 1 2$ can be observed.

\subsubsection{Impurity susceptibility}

For the impurity contribution to the uniform susceptibility
we expect a Curie law, as discussed in Sec.~\ref{sec:obs}.
At tree level, a term of Curie form does only arise from
$\chi_{\rm{u,u}}$, with $T \chi_{\rm{u,u}} = r/8$.
Both $\chi_{\rm{imp},\rm{imp}}$ and $\chi_{\rm{u,imp}}$
are less singular for $r<1$, consistent with $T\chi_{\rm loc}=0$,
see also Appendix~\ref{app:rl}.
Also note that $\chi_{\rm{u,u}}$ is the only of the three terms
where the non-universal number $A_0$ drops out.

We are interested in corrections to $T \chi_{\rm imp}$ to lowest non-trivial
order in $\Ub$, and consequently those corrections can only occur in
$\chi_{\rm{u,u}}$.
A single diagram contributes to first order in $\Ub$ (Fig.~\ref{fig:symdgr1}d),
which gives
\begin{equation}
\Delta \chi_{\rm imp} =
\Ub \frac{V_0^4}{2 A_0^4} \left[
\int_{-\Lambda}^\Lambda dk |k|^r T \sum_n \frac{|\omega_n|^{-2r}}{(i\omega_n-k)^2}
\right]^2
\end{equation}
The $k$ integral can be performed first, with the UV cutoff sent to infinity,
and the frequency summation then leads to
\begin{equation}
\Delta \chi_{\rm imp} =
\frac{1}{T} \left(1-2^{-1-r}\right)^2 \frac{\zeta(1+r)^2}{2\pi^{2(1+r)}} \,
\frac{\Ub T^\epsilon}{A_0^2}
\end{equation}

\begin{figure}
\epsfxsize=3.1in
\centerline{\epsffile{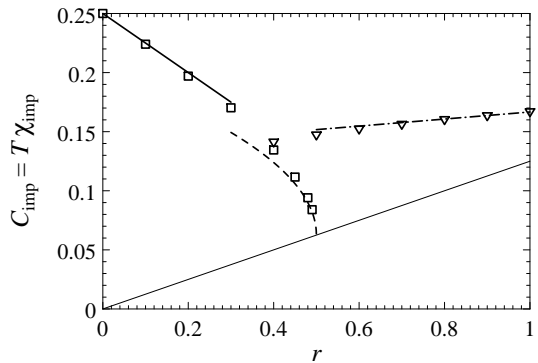}}
\caption{
Numerical data for the impurity susceptibility, $T\chi_{\rm imp}$,
at both the symmetric (squares) and asymmetric (triangles) critical points,
together with the renormalized perturbation theory results from the expansions
in $r$ [Sec.~\protect\ref{sec:weak}, Eq.~(\protect\ref{tchi_weak}), solid],
in $(\frac{1}{2} -r)$ [Sec.~\protect\ref{sec:sym}, Eq.~(\protect\ref{tchi_sym}), dashed], and
in $(1-r)$ [Sec.~\protect\ref{sec:infu}, Eq.~(\protect\ref{tchi_infu}), dash-dot].
The thin line is the value of the SSC fixed point, $T\chi_{\rm imp}=r/8$.
The numerical data are partially taken from Ref.~\protect\onlinecite{GBI};
we have re-calculated the data points near $r=1$, as the logarithmically
slow flow at $r=1$ complicates the data analysis.
}
\label{fig:tchi}
\end{figure}

In the low-energy limit, the combination $(\Ub T^\epsilon/A_0^2)$ approaches
a universal value,
the non-universal prefactors $V_0$ are seen to cancel, and the result has
the expected universal Curie form.
Introducing the renormalized coupling $\U$ we have to
leading order in $\epsilon$:
\begin{equation}
\Delta (T\chi_{\rm imp}) =
\left(1-\frac{1}{2\sqrt{2}}\right)^2 \frac{\zeta(3/2)^2}{2\pi^3} \U \,.
\end{equation}
Using the fixed point value of $\U$ (\ref{u_scr}),
the result for the impurity susceptibility at the SCR fixed point
reads
\begin{equation}
\label{tchi_sym}
T \chi_{\rm imp} = \frac{r}{8} \,+\, 0.1942 \sqrt{\frac{1}{2}-r} \,+\, {\cal O}(\epsilon)
\end{equation}
where $r/8$ is the tree level contribution, and the ${\cal O}(\epsilon)$ represents
higher perturbative terms as above.
This result can be nicely compared to NRG data of Ref.~\onlinecite{GBI},
see Fig.~\ref{fig:tchi}.

\subsubsection{Impurity entropy}

The impurity entropy can be straightforwardly determined from a perturbative
expansion of the impurity part of the thermodynamic potential.
The lowest correction to the tree-level value
$S_{\rm imp} = 2r \ln 2$ is of order $\Ub^2$.
The corresponding contribution to the thermodynamic potential, Fig.~\ref{fig:symdgr1}e,
is given by
\begin{equation}
\Delta \Omega_{\rm imp} = \frac{\Ub^2}{2} \int_0^\beta d\tau \, G_f^2(\tau) G_f^2(-\tau) \,.
\end{equation}
Here $G_f(\tau)$ is the fourier-transformed Green's function $G_f$ (\ref{dressedf})
in the presence of an UV cutoff $\Lambda$.
Taking the temperature derivative we can write the entropy
result as
\begin{equation}
\Delta S_{\rm imp} = \frac{\Ub^2 T^{2\epsilon}}{2 A_0^4} \,
\left[\partial_{T} \int_0^\beta d\tau \,\bar{G}_f^2(\tau) \bar{G}_f^2(-\tau) \,\big|_{r=\frac 1 2} \right]
\end{equation}
where $\bar{G}_f = A_0 G_f$.
The prefactor can be expressed in terms of the renormalized coupling $\U^2$
and will be proportional to $\epsilon$ at the fixed point.
Thus, to leading order the square bracket term can be evaluated at $r=\frac 1 2$,
and is expected to be a universal, finite number in the limit of $\Lambda\to\infty$.
Unfortunately, we were not able to analytically prove the convergence of the integral
as $\Lambda\to\infty$.
We have therefore resorted to a numerical evaluation for finite $\Lambda$ and $T$,
with an extrapolation of the result to $\Lambda/T\to\infty$,
and obtained $[\partial_T ...] = 0.1590 \pm 0.0005$, i.e.,
\begin{equation}
\Delta S_{\rm imp} = 0.159 \frac{\U^2}{2} \,.
\end{equation}
Adding the tree-level result and the $\U^2$ correction we obtain
\begin{equation}
S_{\rm imp} = \ln 2 + (0.03 \pm 0.005) \left(\frac 1 2 - r\right) +\, {\cal O}(\epsilon^{3/2}) \,.
\label{simp_sym}
\end{equation}

\begin{figure}
\epsfxsize=3.1in
\centerline{\epsffile{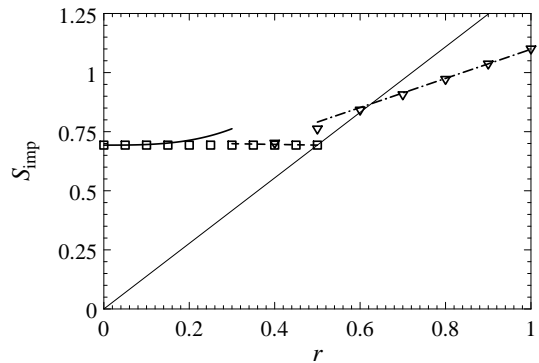}}
\caption{
As Fig. \protect\ref{fig:tchi}, but for the impurity entropy $S_{\rm imp}$;
the NRG values at the SSC fixed point are hardly distinguishable from $\ln 2$.
The perturbative expansions are in
Eq.~(\protect\ref{simp_weak}) (solid),
Eq.~(\protect\ref{simp_sym}) (dashed), and
Eq.~(\protect\ref{simp_infu}) (dash-dot);
the thin line is the value of the SSC fixed point, $T\chi_{\rm imp}=2 r \ln 2$.
}
\label{fig:simp}
\end{figure}

The two expansions for the entropy of the SSC fixed point, (\ref{simp_weak}) and (\ref{simp_sym}),
predict a small positive correction to $\ln 2$ for $0<r<\frac 1 2$, as shown in
Fig.~\ref{fig:simp}, consistent with the notion that entropy should
decrease under RG flow \cite{gtheorem}.
Results from NRG (Ref.~\onlinecite{GBI} as well as ours) show that the deviation
from $\ln 2$ is tiny for all $0<r<\frac 1 2$, which is in principle consistent
with the analytical results.
[Although we identified signatures of a $r^3$ correction to $S_{\rm imp} = \ln 2$
in the small-$r$ regime, the accuracy of our NRG procedure was insufficient
to determine the prefactor \cite{MKMV}.]

\subsubsection{Conduction electron $T$ matrix}

In the Anderson model formulation of the impurity problem,
the conduction electron $T$ matrix is directly proportional
to the physical $f$ electron propagator.
As shown in Appendix~\ref{app:rgsym}, there are no singular propagator
renormalizations in the present problem, $Z=1$ (\ref{zeq1}).
Thus, from Eq.~(\ref{dressedf}) we have the {\em exact} result for the exponent of
the $T$ matrix,
\begin{equation}
\label{T_SCR}
{\rm Im}T(\omega) \propto |\omega|^{-r} \,,
\end{equation}
which holds at the SCR and SCR' fixed points,
and (trivially) also at the SSC fixed point.
A result similar to Eq.~(\ref{T_SCR}) was derived \cite{MVMK} for the SCR fixed
point within the small-$r$ expansion of Sec.~\ref{sec:weak},
in agreement with the notion that both expansions describe the same critical
fixed point.


\section{Maximally particle-hole asymmetric Anderson model}
\label{sec:infu}

In this section we consider a different limit of the pseudogap Anderson
model (\ref{aim}), namely the model with maximal p-h asymmetry.
This means that one of the four possible impurity states will be
excluded: with $\Ub\to\infty$ (keeping $\epsfb$ finite) this is the doubly
occupied one, whereas with $\epsfb\to-\infty$ (keeping $\Ub$ finite) the
empty state is excluded.
As these two situation are related by the p-h transformation (\ref{ph1}),
we will formulate the following for $\Ub\to\infty$.
The Hamiltonian can be written as
\begin{eqnarray}
\label{aiminfu}
{\cal H} &=& \epsfb |\sigma\rangle\langle \sigma|
   + \hybb \left[|\sigma\rangle\langle e| c_\sigma(0) + {\rm h.c.}\right] \\
  &+&\, \int_{-\Lambda}^{\Lambda} d k\,|k|^r \,
  k c_{k\sigma}^\dagger c_{k\sigma}
\nonumber
\end{eqnarray}
where $|\uparrow\rangle$, $|\downarrow\rangle$, and $|{\rm e}\rangle$ represent
the three allowed impurity states.

We shall show that this infinite-$U$ pseudogap Anderson model has a phase
transition between a free-moment and a Kondo-screened phase, which is
accessible to perturbative RG techniques near $r=1$.
In particular, $r=1$ plays the role of the upper-critical dimension.
Furthermore, we will argue in Secs.~\ref{sec:asym} and \ref{sec:AvsK}
that the transitions in both the Anderson and the Kondo model with {\em finite}
p-h asymmetry are in the same universality class, and are described by the
RG presented below, provided that $r>r^\ast$.

\subsection{Trivial fixed points}

For vanishing hybridization $\hybb$, the maximally asymmetric Anderson model features
three trivial fixed points: for $\epsfb<0$ the ground state is the spinful
doublet (LM) with $\ln 2$ entropy.
For $\epsfb>0$ we have an empty-state singlet, which we can identify with
the ASC state of the Kondo model (see below).
The doubly occupied singlet state (labelled ASC') is related to ASC
by the p-h transformation (\ref{ph1}).
For $\epsfb=0$ we have three degenerate impurity states, we refer
to this as the valence-fluctuation fixed point (VFl), with entropy $\ln 3$.
The impurity spin susceptibilities are
\begin{equation}
T\chi_{\rm imp} = \left\{
\begin{array}{ll}
1/4 & \mbox{LM} \\
1/6 & \mbox{VFl} \\
 0  & \mbox{ASC} \\
\end{array}
\right. .
\end{equation}
Again, the conduction electron phase shift is zero at these fixed points
due to the vanishing fixed point value of the hybridization.
The hybridization term, $\hybb$, is irrelevant at the ASC fixed point for
all $r$, and irrelevant at LM for $r>0$.


\subsection{Upper-critical dimension: Expansion around the valence-fluctuation fixed point}
\label{sec:rginfu}

In the following we perform an expansion around the VFl fixed point, i.e.,
around $\epsfb=0$, $\hybb=0$. This will give access to the properties of
the ASC fixed point, i.e., a critical fixed point different from the one
accessed by the RG calculations in Secs.~\ref{sec:weak} and \ref{sec:sym}.

To represent the three impurity states in the infinite-$U$ Anderson model
it is useful to introduce auxiliary fields for pseudo-particles,
$b_s$ (for the empty-state singlet) and $f_\sigma$ (for the spinful doublet).
The required Hilbert space constraint
$b_s^\dagger b_s + f_\sigma^\dagger f_\sigma = \hat{Q} = 1$
will be implemented using a chemical
potential $\lam\to\infty$,
such that observables $\langle\hat{\cal O}\rangle$ have to be calculated
according to \cite{lambda,costi}
\begin{equation}
\langle\hat{\cal O}\rangle =
\lim_{\lam\to\infty}
\frac{\langle\hat{Q}\hat{\cal O}\rangle_{\lam}}{\langle\hat{Q}\rangle_{\lam}} \,,
\label{obs}
\end{equation}
where $\langle\ldots\rangle_{\lam}$ denotes the thermal expectation value
calculated using pseudo-particles in the presence of the chemical potential
$\lam$.
Clearly, in the limit $\lam\to\infty$ the term $\langle\hat{Q}\rangle_{\lam}$
represents the partition function of the physical sector of the Hilbert
space times $\exp(-\lam\beta)$.
As detailed in Ref.~\onlinecite{MKMV}, both numerator {\em and} denominator
of Eq.~(\ref{obs}) have to be expanded in the non-linear couplings to the
required order when calculating observables;
however, the denominator does typically not develop logarithmic singularities
at the marginal dimension.

Furthermore, we need to introduce chemical-potential counter-terms
which cancels the shift of the critical point occurring
in perturbation theory upon taking the limit of infinite UV cutoff.
Technically, this shift arises from the real parts
of the self-energies of the $b_s$ and $f_\sigma$ particles.
We introduce the counter-terms as additional chemical
potential for the auxiliary particles,
\begin{equation}
\delta\lambda_b \, b_s^\dagger b_s \,,~
\delta\lambda_f \, f_\sigma^\dagger f_\sigma \,.
\label{counter}
\end{equation}
The $\delta\lambda_{b,f}$ have to be determined order by order in an expansion in
$\hybb$.
Note that counter-term contributions in observables in general enter both
numerator and denominator in Eq.~(\ref{obs}).

The model (\ref{aiminfu}) can then be written in the
path integral form
\begin{eqnarray}
\label{th}
{\cal S} = \int_0^\beta d \tau
&\bigg[&
   \bar{f}_\sigma (\partial_\tau - \epsfb - \lam - \delta\lambda_f) f_\sigma \nonumber\\
  &+& \bar{b}_s (\partial_\tau - \lam - \delta\lambda_b) b_s \nonumber\\
  &+& \hybb \left(\bar{f}_\sigma b_s c_\sigma(0) + {\rm c.c.}\right) \nonumber\\
  &+& \int_{-\Lambda}^{\Lambda} d k\,|k|^r \,
  {\bar c}_{k\sigma} (\partial_\tau-k) c_{k\sigma}
\bigg]
\end{eqnarray}
where $\lam$ is the chemical potential enforcing the constraint exactly,
and the rest of the notation is as above.
The counter-terms (\ref{counter}) are determined from the real parts of
the self-energies of both the $f_\sigma$ and $b_s$ particles at zero temperature
according to
\begin{eqnarray}
\delta\lambda_f &=& {\rm Re} \Sigma_{f}(\lam+\epsfb + \delta\lambda_f, T=0) \,,\nonumber\\
\delta\lambda_b &=& {\rm Re} \Sigma_{b}(\lam+\delta\lambda_b, T=0)\,,
\end{eqnarray}
note that these real parts diverge linearly with the UV cutoff $\Lambda$.

The model (\ref{th}) shows a transition driven by variation of $\epsfb$
for finite values of $\hybb$.
Tree level scaling analysis shows that
\begin{equation}
{\rm dim}[\hybb] = \frac{1-r}{2} \equiv {\bar r} \,.
\end{equation}
This establishes the role of $r=1$ as upper-critical dimension where
$\hybb$ is marginal.

We now proceed with an RG analysis of (\ref{th}) which will allow
to determine the critical properties for $r\lesssim 1$ -- a brief account
on this appeared in Ref.~\onlinecite{MVLF}.
The RG will treat the auxiliary fields $f_\sigma$ and $b_s$ as usual particles
and consequently determine their propagator renormalizations, anomalous
dimensions etc., but all diagrams are evaluated with $\lam\to\infty$
which ensures the non-trivial character of the expansion.
Renormalized fields and dimensionless couplings are introduced according to
\begin{eqnarray}
f_\sigma &=& \sqrt{Z_f} \, f_{R\sigma} \,, \\
b_s &=& \sqrt{Z_b} \, b_{R} \,, \\
\hybb &=& \frac{\mu^{\bar{r}} Z_\hyb}{\sqrt{Z_f Z_b}} \hyb
\,. \label{gdef}
\end{eqnarray}
No renormalizations are needed for the bulk fermions as their self-interaction
is assumed to be irrelevant in the RG sense.

In contrast to the field theories analyzed in Secs.~\ref{sec:rgweak} and
\ref{sec:SSCRG} (where the non-linear coupling is used to tune the system
through the phase transition),
the theory (\ref{th}) contains two parameters, namely the tuning
parameter $\epsfb$ and the non-linear coupling $\hybb$.
The RG is conveniently performed {\em at} criticality, i.e.,
we assume that $\epsfb$ is tuned to the critical line,
and RG is done for the coupling $\hyb$ --
this naturally results in an infrared stable fixed point.
To two-loop order we obtain the following RG beta function:
\begin{eqnarray}
\beta(\hyb) &=& - {\bar r} \hyb + \frac{3}{2} \hyb^3 + 3 \hyb^5 \,
\label{betaginfu}
\end{eqnarray}
with the calculation given in Appendix~\ref{app:rginfu}.
Generally, the higher-order corrections to the one-loop result
appear to be small in the present expansion.
One can also consider the flow away from criticality, i.e.,
the flow of the tuning parameter $\epsf$, either using $S^2$
insertions in the field-theoretic formulation or explicitly
within momentum-shell RG.
The resulting correlation length exponent is in Eq.~(\ref{nuz_infu})
below.

The structure of the above RG is very similar to the one of the
$(4-\epsilon)$ expansion of the $\phi^4$ model for magnets, where
the mass term drives the transition, and the non-linear coupling
has a non-trivial stable fixed point at criticality below four
dimensions.
Thus, the fixed point with finite $\hyb$ (\ref{fp}) corresponds
to the Wilson-Fisher fixed point, whereas $\hyb=0$ is the analogue
of the Gaussian fixed point in the magnetic context, see
Fig.~\ref{fig:flowinfu}.
The parameters $\hyb$ and $\epsf$ play the role of the interaction
and the mass, respectively.

\subsection{$r^\ast < r < 1$}

For $r<1$ the trivial fixed point $\hyb^\ast=0$ is unstable,
and the critical properties are instead controlled by an interacting
fixed point at
\begin{equation}
{\hyb^\ast}^2 = \frac{2}{3} \bar r - \frac 8 9 {\bar r}^2 \,.
\label{fp}
\end{equation}
At this asymmetric critical fixed point (ACR), we find anomalous field
dimensions
$\eta_b = 2 {\hyb^\ast}^2 + 2 {\hyb^\ast}^4$, $\eta_f = {\hyb^\ast}^2 + 2 {\hyb^\ast}^4$.
The resulting RG flow diagram is shown in Fig.~\ref{fig:flowinfu}a.
The ACR fixed point (Fig. \ref{fig:flowinfu}b) shifts to larger values of
${\hyb^\ast}^2$, $|\epsf^\ast|$ with decreasing $r$, and the expansion
can be expected to break down for small $r$.
The numerical results of Ref.~\onlinecite{GBI} show that this is
the case at $r^\ast\approx 0.375$, where ACR merges with SCR, and
p-h symmetry is dynamically restored.
Then, Eq. (\ref{th}) together with an expansion in $\hyb$, $\bar{r}$
yields the correct description of the critical properties for
$0.375\approx r^\ast < r < 1$.
Using NRG we have numerically confirmed this expectation, i.e., the properties of
the critical fixed point of the model (\ref{th}) vary continuously as function
of $r$ for $r^\ast < r < 1$.


\subsection{$r=1$}

For all $r\geq 1$, a phase transition still occurs in the asymmetric
Anderson and Kondo models, but it is controlled by the non-interacting
VFl fixed point at $\hyb=\epsf=0$.
For the marginal case $r=1$, i.e., at the upper-critical dimension, we
expect logarithmic flow.
In the following we explicitly keep the UV cutoff $\Lambda$,
and discuss RG under cutoff reduction $\Lambda \to \lambda\Lambda$.
We restrict ourselves to criticality, where the RG beta function
to one-loop order is
\begin{eqnarray}
\beta(\hyb) \equiv \frac{d \hyb}{d \ln \lambda} =
\frac{3}{2} \hyb^3
\end{eqnarray}
i.e., the hybridization is marginally irrelevant.
The RG equation can be integrated to give
\begin{eqnarray}
\label{hybr1}
\hyb^2(\lambda) = \frac{\hybb^2}{1-3\hybb^2\ln\lambda}
\end{eqnarray}
with $\hyb(\lambda=1) = \hybb$.
This result will be used below to determine logarithmic corrections
for a number of observables.


\subsection{$r>1$}

For bath exponents $r\!>\!1$ the coupling $\hybb$ in the theory (\ref{th})
is irrelevant in the RG sense.
The critical system flows to the VFl fixed point, Fig.~\ref{fig:flowinfu}b,
and the transition becomes a level crossing with perturbative corrections.

Observables can then be obtained by straightforward perturbation theory.
Consider, e.g., the boson self-energy (Fig. \ref{fig:infudgr2}c below):
\begin{eqnarray}
\Sigma_{b}(i\nu_n) = \hybb^2\, T \sum_{n}
\int \frac{d k\,|k|^r}{i\omega_n\!-\!k} \,
\frac{1}{i\nu_n \!-\! i\omega_n \!-\!\lam\!-\!\epsfb} \,,
\nonumber
\end{eqnarray}
the expression for the fermion self-energy is similar.
At the transition, $\epsfb\!=\!0$, the self-energies show threshold behavior at $T\!=\!0$,
$-{\rm Im}\,\Sigma_{f}(\bar{\omega}+i\eta)/\pi \propto \hybb^2 \bar{\omega}^r \Theta(\bar{\omega})$
with $\bar{\omega} = \omega-\lam$.
The low-energy behavior of the $f_\sigma$ propagator follows as
\begin{eqnarray}
-{\rm Im} \, G_{f}(\bar{\omega}+i\eta)/\pi =
(1-A) \delta(\bar{\omega}) + B |\bar{\omega}|^{r-2} \Theta(\bar{\omega})
\label{gf_pt}
\end{eqnarray}
with $A,B \propto \hybb^2$.
The $b_s$ propagator has a similar form -- we will use these results
below to explicitly calculate the local susceptibility.

\subsection{Observables near criticality}
\label{sec:infuobs}

We start with the correlation length exponent, $\nu$, of the asymmetric
critical fixed point.
In the field-theoretic RG scheme it has to be determined via composite operator
insertions into the action (\ref{th}), which take the role of mass terms
driving the system away from criticality.
The lowest-order result for $\nu$ is
\begin{equation}
\frac{1}{\nu} = r + {\cal O}(\bar{r}^2) ~~~~(r<1) \,,
\label{nuz_infu}
\end{equation}
with details of the derivation given in Appendix~\ref{app:rginfu}.
For $r\geq 1$ the transition is a level crossing, formally $\nu=1$.

In the calculation of observables like susceptibilities etc.
the UV behavior depends on whether we are above or below the
upper-critical dimension $r=1$.
For $r\geq 1$ the cutoff $\Lambda$ has to be kept explicitly, as integrals
will be UV divergent.
This also implies that hyperscaling is violated, and no $\omega/T$ scaling
in dynamics occurs, as usual for a theory above the upper-critical dimension.

For $r<1$ the UV cutoff $\Lambda$ can be sent to infinity
after taking into account the contributions of the counter-terms (\ref{counter}),
as the remaining integrals are UV convergent.
This is in contrast to the expansions for $r\gtrsim 0$ (Sec.~\ref{sec:weak})
and for $r\lesssim 1/2$ (Sec.~\ref{sec:sym}), which are effectively
both {\em above} a {\em lower}-critical dimension, and where intermediate quantities can
display UV divergencies.
Notably, for all $r<1$ expansions the low-energy observables are fully universal, i.e.,
cutoff-independent, and hyperscaling is fulfilled.

\subsubsection{Local susceptibility}

The anomalous exponent $\eta_\chi$ associated with the local susceptibility
is calculated as above
by determining the $\chi_{\rm loc}$ renormalization
factor, $Z_\chi$, using minimal subtraction of poles,
and then employing
\begin{equation}
\label{etachidef2}
\eta_\chi = \beta(\hyb) \frac{d \ln Z_\chi}{d \hyb} \bigg|_{\hyb^\ast} \,.
\end{equation}
Here we have $\chi_{\rm loc} = 1/\omega$ at tree level,
in contrast to the p-h symmetric problem of Sec.~\ref{sec:sym}.

\begin{figure}[!t]
\epsfxsize=3.2in
\centerline{\epsffile{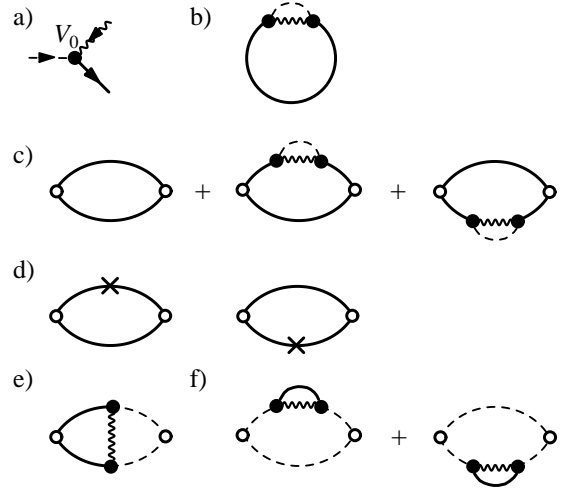}}
\caption{
Feynman diagrams for the infinite-$U$ Anderson model.
Full/wiggly/dashed lines denote $f_\sigma$/$b_s$/$c_\sigma$ propagators,
the cross is the counter-term (\protect\ref{counter}).
a) Bare interaction vertex $\hybb$.
b) $\hybb^2$ contribution to the partition function, which also appears
in the denominator of Eq.~(\protect\ref{obs}).
c),d) Diagrams entering the local susceptibility, $\chi_{\rm imp,imp}$, to order $\hybb^2$.
e) $\hybb^2$ contribution to $\chi_{\rm u,imp}$.
f) $\hybb^2$ contributions to $\chi_{\rm u,u}$.
}
\label{fig:infudgr1}
\end{figure}

The leading contributions to $Z_\chi$ arise from the diagrams
in Fig.~\ref{fig:infudgr1}c, see Appendix~\ref{app:rginfu} for
details.
From the $Z_\chi$ expression
\begin{equation}
Z_\chi = 1- \frac{\hyb^2}{\bar{r}}
\end{equation}
we find $\eta_\chi$ in an expansion in $\bar{r}$,
\begin{equation}
\eta_\chi =  \frac{2}{3}\,(1-r) + {\cal O}(\bar{r}^2) ~~~~(r<1) \,.
\label{etachi_infu}
\end{equation}
This result is again in good agreement with numerical findings \cite{insi},
with a comparison given in Fig.~\ref{fig:etachi}.

We now describe the evaluation of the logarithmic corrections present
at $r=1$.
Using the lowest-order results
$\beta(\hyb) = \frac{3}{2}\hyb^3$ and $\eta_\chi(\hyb)=2\hyb^2$
we can integrate Eq.~(\ref{etachidef2}) to yield
\begin{equation}
\label{log2}
Z_\chi = \left(\frac{\hyb(\lambda)}{\hybb}\right)^{4/3} \,.
\end{equation}
The explicit scaling relation \cite{zj} for $\chi_{\rm loc}$ at $T=0$ reads
\begin{equation}
\chi_{\rm loc}(i\omega,\hybb,\Lambda) =
Z_\chi \, (\lambda\Lambda)^{-1}
\chi_{\rm loc}\left(\frac{i\omega}{\lambda\Lambda},\hyb(\lambda),\Lambda=1\right) \,.
\end{equation}
To analyze the frequency dependence of $\chi_{\rm loc}$ we choose
$\lambda = \lambda^\ast = \omega/\Lambda$ and employ
(\ref{log2}) together with (\ref{hybr1}) to express the renormalization
factor in terms of $\lambda$. This leads to
\begin{eqnarray}
\chi_{\rm loc}(i\omega,\hybb,\Lambda) =
\frac{\omega^{-1}}{
\left(
1-3\hybb^2\ln\frac{\omega}{\Lambda}
\right)^{2/3}} \,
\chi_{\rm loc}(i,\hyb(\lambda),1) \,.
\nonumber
\end{eqnarray}
To obtain the leading (i.e. multiplicative) logarithms, the last term can be
approximated by its fixed point value, which simply gives a constant --
this neglects sub-leading additive logarithmic corrections.
We finally obtain the result
\begin{equation}
\chi_{\rm loc}(\omega) \propto \frac{1}{\omega |\ln\omega|^{2/3}} ~~~~(r=1) \,.
\end{equation}
valid at criticality for $\omega \ll \Lambda$.
Note that the structure of the logarithms in our problem is different
from e.g. that of the Hertz-Millis theory at the upper-critical dimension studied
in Ref.~\onlinecite{pankov}.
In our problem both $\hyb$ and $Z_\chi$ flow to zero according to
Eq.~(\ref{log2}); whereas in Ref.~\onlinecite{pankov} the renormalization factor $Z$
tends to a non-universal constant as $\lambda\to 0$ -- this
leads to the absence of multiplicative logarithms.

Above the upper-critical dimension, $r>1$, we have simply $\eta_\chi =0$ and thus
$\chi_{\rm loc} \propto 1/T$ or $\propto 1/\omega$.
We shall explicitly demonstrate the calculation of the local impurity
susceptibility using bare perturbation theory.
To lowest non-trivial order, $\chi_{\rm loc}$ is given by the convolution
of two $f_\sigma$ propagators (\ref{gf_pt}),
calculated with the self-energy to second order in $\hybb$.
Note that no vertex corrections occur to this order due to the
structure of the interaction.
When calculating $\chi_{\rm loc}$ the $T\to 0$ limit has to be taken with
care, as the exponentially small tail in ${\rm Im} G_{f}(\bar{\omega})$ at
$\bar{\omega}<0$ contributes to $\chi_{\rm loc}$.
One obtains for the imaginary part $\chi''_{\rm loc}$ for $T\to 0$ and $1<r<2$:
\begin{equation}
\chi''_{\rm loc}(\omega)/\pi = \frac{1-2A}{6} \, \frac{\delta(\omega)\omega}{T} +
\frac{B}{3} \, |\omega|^{r-2} {\rm sgn}(\omega) \,.
\end{equation}
This shows that $\omega/T$ scaling is violated, as the finite frequency part
obeys $\chi_{\rm loc}(\omega\gg T) \propto \omega^{r-2}$, but
$\chi_{\rm loc}(\omega\ll T)\propto T^{-1}$.

Moving away from criticality, one of the $f_\sigma$, $b_s$ propagators looses
its $\delta(\omega-\lambda_0)$ contribution --
for $\epsf>0$ ($t>0$) this is $G_{f_\sigma}$, indicating that free-moment
behavior is absent in this regime.
Consequently, the zero-temperature static local susceptibility is finite,
but diverges upon approaching the critical point according to
$\chi_{\rm loc} \propto t^{r-2}$.
For $t<0$, $\chi_{\rm loc} \propto 1/T$.
Thus, the order parameter $m_{\rm loc}$ jumps at the transition point
for $r>1$.

\subsubsection{Impurity susceptibility}

The evaluation of the impurity susceptibility requires the summation
of the diagrams in Fig.~\ref{fig:infudgr1}c--f, appearing in the numerator of the
corresponding Eq.~(\ref{obs}), and also a careful treatment of the
denominator, as we are interested in
terms being non-singular as function of $\bar{r}$.

The diagrams are conveniently evaluated in imaginary time, e.g.,
the first correction to $\chi_{\rm imp,imp}$ in Fig.~\ref{fig:infudgr1}c
gives
\begin{equation}
\frac{1}{2} e^{-\lam\beta} \hybb^2
\int_0^\beta d\tau \int_0^\tau d\tau_2 \int_0^{\tau_2} \! d\tau_1
G_{c0}(\tau_2-\tau_1)
\end{equation}
where the limit $\lam\to\infty$ has been taken in the $f_\sigma$ and $b_s$
propagators.
$G_{c0}$ is the conduction electron Green's function at the impurity site,
which contains a momentum integral.
The other diagrams in Fig.~\ref{fig:infudgr1} can be written down similarly.
Performing the $\tau$ integrals first, one obtains
\begin{eqnarray}
{\rm(8c)} &=&
\frac{1}{T} - \hybb^2 \int_0^\Lambda \!\!dk \frac{k^r}{k^3} \left[
\frac{2k}{T} + \left(4+\frac{k^2}{T^2}\right) \tanh \frac{k}{2T} \right] ,
\nonumber \\
{\rm (8d)} &=&
- \frac{\hybb^2 \Lambda}{T^2} \,,
\nonumber \\
{\rm (8e)} &=& \hybb^2 \int_0^\Lambda \!\!dk k^r
\frac{2[3+\cosh(k/T)]k/T - 4\sinh(k/T)}{k^3 [1+\cosh(k/T)] },
\nonumber \\
{\rm (8f)} &=& \hybb^2 \int_0^\Lambda \!\!dk \frac{k^r}{k^3} \bigg[
4 \tanh \frac{k}{2T} - \nonumber\\
&&~~~~~- \left(\frac{2k}{T}+ \frac{k^2}{T^2} \tanh \frac{k}{2T}\right) \cosh^{-2}\!\frac{k}{2T}
\bigg] ,
\end{eqnarray}
where all terms have to be multiplied by $e^{-\lam\beta}/2$.
Fig.~\ref{fig:infudgr1}d are the contributions from the counter-terms (\ref{counter}),
which evaluate to $\delta\lambda_b = 2\delta\lambda_f = 2\hybb^2\Lambda$.
The denominator, $\langle\hat{Q}\rangle_{\lam}$,
being $3 e^{-\lam\beta}$ to zeroth order in $V_0$,
receives corrections from the diagram in Fig.~\ref{fig:infudgr1}b
and from the counter-terms (\ref{counter}), resulting in
\begin{eqnarray}
\langle\hat{Q}\rangle_{\lam} =
3 - 4 \frac{\hybb^2\Lambda}{T} + 4 \hybb^2 \int_0^\Lambda \!\!dk \frac{k^r}{kT} \tanh\frac{k}{2T} \,,
\end{eqnarray}
to be multiplied with $e^{-\lam\beta}$.
Now we are in the position to collect all contributions
to $\chi_{\rm imp}$ to second order in $\hybb$:
\begin{eqnarray}
\Delta\chi_{\rm imp} &=& - \frac{\hybb^2 T^{-\bar{r}}}{6T}
\bigg[
2 \int_0^{\Lambda/T} \frac{d x\,x^r}{x^3}\, \frac{\sinh x - x}{1+\cosh x} \\
&& ~~~~~~~~~~+
\int_0^{\Lambda/T} \frac{d x\,x^r}{3x} \left(\frac{\sinh x}{1+\cosh x} - 1\right)
\bigg] . \nonumber
\end{eqnarray}
Note that intermediate terms of the form $T^{-2}$ have cancelled in numerator
and denominator of Eq.~(\ref{obs}).
For $r<1$ the momentum integrals are UV convergent, and do not develop poles in $\bar{r}$,
i.e., the poles present in the $\chi_{\rm imp,imp}$ diagrams have been cancelled
by contributions from Fig.~\ref{fig:infudgr1}e/f.
In the low-energy limit, the product $\hybb^2 T^{-\bar{r}}$ approaches a
universal value. Thus $\chi_{\rm imp}$ has indeed Curie form,
with a fully universal prefactor depending on $r$ only.
For $r>1$ the integrals require an explicit UV cutoff, but no correction
to the Curie term arises, as $(\hybb^2 T^{-\bar{r}})$ vanishes
as $T\to 0$.

Performing the integral for $r<1$ and expressing the result in terms of the
renormalized coupling $\hyb$,
the impurity susceptibility reads
\begin{eqnarray}
T \chi_{\rm imp}
= \frac{1}{6} - \left(\frac{1}{6} - \frac{\ln 2}{9}\right) \hyb^2 + {\cal O}(\hyb^4)
\end{eqnarray}
With the value of the coupling at the ACR fixed point (\ref{fp})
we finally find, to leading order in $(1-r)$,
\begin{equation}
\label{tchi_infu}
T \chi_{\rm imp} =
\left\{
\begin{array}{ll}
\frac{1}{6} - 0.02988 (1-r) + {\cal O}(\bar{r}^2)   & (r < 1) \\[2mm]
\frac{1}{6}                                         & (r \geq 1) \\
\end{array}
\right.
\,,
\end{equation}
to be compared with the numerical results in Fig.~\ref{fig:tchi}.

\subsubsection{Impurity entropy}
\label{sec:infuentr}

The impurity contribution to the entropy can be derived from the
free energy as above. At the VFl fixed point the entropy is
$S_{\rm imp} = \ln 3$, and the lowest-order correction is
computed by expanding the free energy in $\hyb$.
Note that this correction vanishes for $r\geq 1$,
as $\hyb^\ast=0$ there.

The calculation of the impurity entropy in the presence of a constraint for
pseudoparticles has been discussed in Appendix C of Ref.~\onlinecite{MKMV}.
The limit $\lam\to\infty$ suppresses all contributions from the unphysical
part of the Hilbert space, in particular disconnected diagrams in the partition
function. Remarkably, this leads to the appearance of disconnected
diagrams in higher-order terms of the expansion for the thermodynamic
potential $\Omega$.
The second-order diagram for $\Omega$, shown in Fig.~\ref{fig:infudgr1}b,
evaluates to
\begin{eqnarray}
\label{omimp3}
\Delta \Omega_{\rm imp} =
\frac{2\hybb^2}{3} \int_{-\Lambda}^{\Lambda} d k \, \frac{|k|^r}{k} \tanh \frac{k}{2T} \,.
\end{eqnarray}
There is also a contribution to $\Omega_{\rm imp}$ from the counter-terms,
but this is temperature-independent and does not contribute to the entropy.
Taking the temperature derivative we find:
\begin{eqnarray}
\Delta S_{\rm imp} =
- \frac{\hybb^2}{3} T^{-2\bar{r}}
\int_{-\Lambda/T}^{\Lambda/T} d x\, |x|^r \cosh^{-2} \frac{x}{2} \,.
\end{eqnarray}
This integral can be performed in the limit of infinite UV cutoff.
[In contrast, the integral in (\ref{omimp3}) is UV divergent, as discussed in
Sec.~{\ref{sec:obs}.)
Expressing the result in terms of the renormalized hybridization
and taking the limit $r\to 1$ we have
\begin{eqnarray}
\label{simp3}
S_{\rm imp} = \ln 3 - \frac{8\ln 2}{3} \,\hyb^2 \,.
\end{eqnarray}
As expected, the entropy correction is fully universal and finite in the
limit $T\to 0$.
Inserting the fixed point value of the coupling $\hyb$ into Eq.~(\ref{simp3}),
we find the impurity entropy as
\begin{equation}
\label{simp_infu}
S_{\rm imp} =
\left\{
\begin{array}{ll}
\ln 3 - \frac{8 \ln 2}{9} (1-r) + {\cal O}(\bar{r}^2)   & (r < 1) \\
\ln 3                                                   & (r \geq 1) \\
\end{array}
\right.
\,.
\end{equation}
NRG calculations\cite{GBI} have determined $S_{\rm imp}$ for $r$ values below unity;
a comparison is shown in Fig.~\ref{fig:simp}.
As with most of the observables obtained within the $(1-r)$ expansion the agreement of
the lowest-order result with numerics is surprisingly good even for $r$ values well away
from unity, indicating that higher-loop corrections are small.

The results for the RG flow and the impurity entropy have an interesting corollary:
Stability analysis shows that for $r^\ast < r < \frac 1 2$ the RG flow at criticality
and small p-h asymmetry is from SCR to ACR.
The entropy of SCR approaches $\ln 2$ as $r\to\frac 1 2$.
Both NRG and the above expansion indicate that the entropy of ACR is {\em larger}
than $\ln 2$ for $r \lesssim \frac 1 2$, Fig.~\ref{fig:simp}.
Thus we have $S_{\rm imp, ACR} > S_{\rm imp, SCR}$ for $r^\ast < r < \frac 1 2$,
i.e., the impurity part of the entropy increases under RG flow
(see also Fig.~\ref{fig:snrg} below),
in contradiction to the so-called $g$-theorem \cite{gtheorem}!
(This is not a fundamental problem, as the present model has effectively
long-ranged interactions, and is not conformally invariant, such that the
proof of the $g$-theorem does not apply.)

\subsubsection{Conduction electron $T$ matrix}

The $T$ matrix in the Anderson model is given by
$T(\omega) = \hybb^2 G_f(\omega)$.
The physical $f$ propagator $G_f$ is a convolution of the
auxiliary $f_\sigma$ and $b_s$ propagators, i.e.,
the propagator of the composite operator $(f_\sigma^\dagger b_s)$.
The anomalous exponent is obtained from
\begin{equation}
\label{etaTdef}
\eta_T = \beta(\hyb) \frac{d \ln Z_T}{d \hyb} \bigg|_{\hyb^\ast} \,.
\end{equation}

As in Ref.~\onlinecite{MVMK} we are able to determine an {\em exact} result for
the anomalous exponent, valid to all orders in perturbation theory.
The argument is based on the diagrammatic structure of the $T$ matrix,
namely the relevant diagrams can be completely constructed from full $\hyb$
interaction vertices and full $f/b$ propagators \cite{MVMK}.
This leads to the relation between $Z$ factors
\begin{equation}
Z_T^{-1} = \frac{Z_\hyb^2}{Z_f Z_b} \,.
\end{equation}
This equation can be plugged into (\ref{gdef}).
Taking the logarithmic derivative at fixed bare coupling
and using $\beta(\hyb)/\hyb = 0$ at any fixed point with finite $\hyb^\ast$,
one obtains the exact result
\begin{equation}
\eta_T= 2\bar{r} ~~~\Rightarrow~~~
{\rm Im} T(\omega) \propto |\omega|^{-r}.
\label{exact}
\end{equation}
Whereas Eq. (\ref{exact}) applies to the p-h asymmetric fixed point (ACR),
and is valid for $r^\ast<r<1$, the results of Sec.~\ref{sec:symobs}
and Ref.~\onlinecite{MVMK} have established the same critical behavior for
the $T$ matrix for the symmetric fixed point (SCR) for $0<r<\frac{1}{2}$.
Thus, we conclude that all critical fixed points for $0<r<1$ in the
pseudogap Anderson and Kondo models display a $T$ matrix behavior of
${\rm Im} T(\omega) \propto |\omega|^{-r}$.

The logarithmic correction to the $T$ matrix at $r=1$ are evaluated in a manner
similar to the one for the local susceptibility above.
With $\beta(\hyb) = \frac{3}{2}\hyb^3$ and $\eta_T(\hyb)=3\hyb^2$ --
note that $\eta_T = \eta_f+\eta_b$ only holds at one-loop level
because $Z_\hyb=1$ at this order --
we can integrate (\ref{etaTdef}) to find
\begin{equation}
\label{log4}
Z_T = \left(\frac{\hyb(\lambda)}{\hybb}\right)^2 \,.
\end{equation}
With the general scaling relation
\begin{equation}
T(i\omega,\hybb,\Lambda) =
Z_T \, (\lambda\Lambda)^{-1}
T\left(\frac{i\omega}{\lambda\Lambda},\hyb(\lambda),\Lambda=1\right)
\end{equation}
and (\ref{hybr1}) we find
\begin{eqnarray}
T(i\omega,\hybb,\Lambda) =
\frac{\omega^{-1}}{
1-3\hybb^2\ln\frac{\omega}{\Lambda}}
\,
T(i,\hyb(\lambda),1) \,.
\nonumber
\end{eqnarray}
As above, the leading logarithms are of multiplicative character,
with the final result
$T(\omega) \propto 1/(\omega |\ln\omega|)$.
This implies for the $T$ matrix spectral density the behavior
\begin{equation}
{\rm Im} T(\omega) \propto \frac{1}{\omega |\ln\omega|^2} ~~~~(r=1) \,.
\end{equation}

Above the upper-critical dimension, $r>1$, we have again $\eta_T =0$
and ${\rm Im} T(\omega) \propto \delta(\omega)$.


\section{General particle-hole asymmetry}
\label{sec:asym}

Here we comment on the general case of finite p-h asymmetry.
Starting with the trivial fixed points, tree-level power counting shows
that LM is stable w.r.t. p-h asymmetry, with a scaling dimension of $-r$.
In contrast, SSC is unstable towards ASC, and p-h asymmetry grows near SSC
with a scaling dimension of $r$.
Finally, at the free-impurity fixed point (FImp) p-h asymmetry grows
under RG with a scaling dimension of unity and the systems flow towards
VFl.
VFl itself is stable w.r.t. a deviation from maximal p-h asymmetry.

In the $r$ ranges where the RG expansions of this paper are perturbatively
controlled, we can immediately come to conclusions about the stability of
the critical fixed points: SCR will be stable w.r.t. finite p-h asymmetry for
$r$ close to 0, but unstable for $r$ close to $\frac 1 2$.
Similarly, for $r$ close to or larger than unity,
ACR is stable w.r.t. a deviation from maximal p-h asymmetry,
in other words it is safe to discard one of the two impurity states,
$|{\rm d}\rangle$ or $|{\rm e}\rangle$, for the discussion of the critical properties, and
to work with the perturbative expansion of Sec.~\ref{sec:infu}.
One has to keep in mind that ACR moves towards smaller effective p-h asymmetry
(larger values of $|\epsf^\ast|$) upon decreasing $r$,
and the perturbative expansion around VFl breaks down as
$r\to {r^\ast}^+$ where p-h symmetry is dynamically restored.

The numerics of Ref.~\onlinecite{GBI} gives no indications for additional
fixed points in the case of finite p-h asymmetry which would not be present
in the maximally asymmetric model;
our RG results are consistent with this.
Thus, the critical properties of a pseudogap Kondo or Anderson model with general
p-h asymmetry are always identical to the ones of the maximally asymmetric
model of Sec.~\ref{sec:infu}.

\begin{figure}[t]
\epsfxsize=3.3in
\centerline{\epsffile{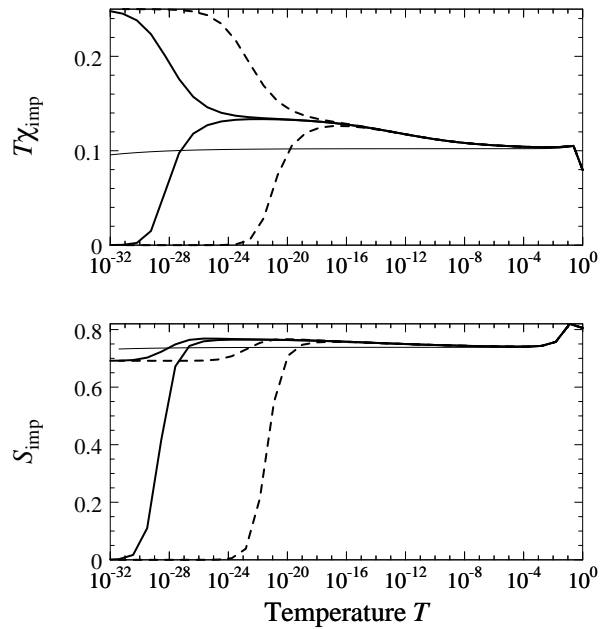}}
\caption{
NRG results for a slightly p-h asymmetric Anderson model at $r=0.45$ close
to criticality (ACR).
Parameter values are a hybridization strength $\pi V_0^2 \rho(0) = 1$,
$\epsfb = -0.5$, and $\Ub =$ 0.9949, 0.99493 (dashed), 0.9949213, 0.9949215 (solid).
The thin lines are for a p-h symmetric model very close to criticality (SCR),
with $\epsfb = -0.487$, $\Ub = 0.974$.
The NRG runs were done using a discretization parameter of $\Lambda = 9$,
keeping $N_s = 650$ levels.
The two-stage flow described in the text can be clearly seen,
i.e., there is a small energy scale $T^\ast$ where the system flows from
ACR to ASC or LM (solid: $T^\ast \approx 10^{-28}$, dashed: $T^\ast \approx 10^{-22}$),
and a larger scale $T_{\rm ACR} \approx 10^{-12}$ where the system
flows from SCR to ACR.
a) Impurity susceptibility $T\chi_{\rm imp}$.
b) Impurity entropy $S_{\rm imp}$.
The behavior of the entropy illustrates the point made in Sec.\protect\ref{sec:infuentr}:
$S_{\rm imp}(T)$ decreases as function of $T$, i.e., {\em increases} along the RG flow,
for $T^\ast< T < T_{\rm ACR}$ due to $S_{\rm imp, ACR} > S_{\rm imp, SCR}$.
This ``uphill flow'' does not violate thermodynamic stability criteria,
as the total entropy (impurity plöus bath) of the system still decreases
under RG.
}
\label{fig:snrg}
\end{figure}

For the RG treatment of a model with general p-h asymmetry close to criticality
one has to envision a {\em two-step} RG procedure, as usual in problems with
different energy scales.
Suppose we start near criticality from small hybridization and small p-h
asymmetry.
For $r > r^\ast$, the scaling dimension of p-h symmetry breaking term is
largest: at tree level we have ${\rm dim}[2\epsfb+\Ub] = 1$ and
${\rm dim}[\hybb] = (1-r)/2$.
Thus the initial model parameters flow towards large p-h asymmetry first.
This flow drives the system into the regime described by the maximally
p-h asymmetric Anderson model of Sec.~\ref{sec:infu}, thereby renormalizing
the parameters $\epsf$ and $\hyb$. Then, the RG of Sec.~\ref{sec:infu} takes
over, with a flow towards ACR, and finally to one of the two stable phases
(if the system is not exactly at criticality).
Thus, a low-energy scale $T^\ast$ and a higher scale $T_{\rm ACR}$ exist,
where $T_{\rm ACR}$ characterizes the approach of the effective p-h asymmetry
towards the ACR fixed point.
Note that for $r>2$ the fastest flow is the one of $\hyb$ to zero, i.e., the system
quickly approaches the VFl fixed point.
The described behavior is nicely borne out by NRG calculations for the Anderson model,
see Fig.~\ref{fig:snrg}.

Finally, for $r<r^\ast$ the initial flow is dominated by the decrease of
p-h asymmetry -- this cannot be captured by our Anderson model RG, but is contained
in the Kondo treatment -- and the system approaches SCR, before it finally departs
to one of the two stable phases.
If the system is on the strong-coupling side of the transition ($t>0$), then
the behavior near SCR is multicritical (see also Fig. 16a
of Ref.~\onlinecite{GBI}):
two low-energy scales exist which describe the departure of the flow from SCR,
namely $T^\ast$ (for the deviation of $j$ from $j^\ast$) and a {\em lower} scale
$T_{\rm ASC}$ (for the subsequent growth of p-h asymmetry when flowing towards the
ASC fixed point).

We note that a recent investigation of the pseudogap Anderson
model \cite{logan03} using the local-moment approach \cite{LMA}
has found indications of a line of critical fixed points
in the p-h asymmetric case, parametrized by p-h asymmetry.
We believe that this is an artifact of the employed approximation
scheme, as (i) NRG calculations strongly hint towards a single asymmetric critical fixed
point (for fixed $r$), i.e., the fixed-point level spectrum at criticality
does not depend on the initial p-h asymmetry (e.g. the ratio $\Ub/\epsfb$),
and (ii) our analytical RG shows that, at least near $r=1$, the
scaling dimension of p-h asymmetry is largest, and the sketched two-step
RG (which directly leads to consider an infinite-$U$ Anderson model to
describe the critical behavior) is a controlled approach for arbitrary initial
p-h asymmetry.


\section{Relation between Anderson and Kondo models}
\label{sec:AvsK}

This section shall highlight the relation between the pseudogap
Anderson and Kondo impurity models.

On the one hand, it is well-known that the Anderson model reduces to the
Kondo model in the so-called Kondo limit, see Sec.~\ref{sec:models}.
This mapping covers the far left-hand side of the flow diagrams in
Figs.~\ref{fig:flowsym} and \ref{fig:flowinfu}, and suggests that the
phase transition at small $r$ in the Anderson model in this Kondo limit
is described by the Kondo RG of Sec.~\ref{sec:weak}.

On the other hand, we have argued that the flow of the Kondo model
can be naturally understood in terms of the variables of the Anderson
model.
In particular, the RG expansions of Secs.~\ref{sec:sym} and \ref{sec:infu},
describing transitions of the Anderson model near $r=\frac{1}{2}$ and $r=1$,
also apply to the Kondo model.
Clearly, in thinking about this ``mapping'' of the Kondo to the Anderson
model one has to view the Anderson model as an effective {\em low-energy}
theory of the Kondo model.
The impurity states of this effective Anderson model are thus many-body states
obtained after integrating out high-energy degrees of freedom from the Kondo model,
i.e., dressed impurity states.

A formal way to obtain an Anderson model from a Kondo model is
a strong-coupling expansion:
The $|\uparrow\rangle$ and $|\downarrow\rangle$ states of the Anderson model
are bare impurity states, whereas $|{\rm e}\rangle$ and $|{\rm d}\rangle$ represent the
impurity with either a hole or an electron of opposite spin tightly bound to the
impurity.

The above can also be made plausible in a slightly different way,
in the following for a p-h asymmetric situation:
Let us consider the Kondo problem with a hard-gap DOS \cite{hardgap}:
in the presence of p-h asymmetry it shows a first-order quantum transition, i.e.,
a level crossing, between a Kondo-screened singlet and a spin-$\frac{1}{2}$ doublet
state.
As we can understand the asymmetric pseudogap DOS as consisting
of an asymmetric high-energy part and a (asymptotically) symmetric
low-energy part, we can obtain an effective theory for the pseudogap Kondo model
by coupling the above mentioned three (many-body) impurity states,
obtained by integrating out high-energy degrees of freedom from the band,
to the remaining low-energy part of the conduction electron spectrum.
A crucial ingredient is now the p-h asymmetry of the original model.
It is clear that upon integrating out the high-energy part of the bath
{\em two} many-body singlet states arise, namely $|{\rm e}\rangle$ and $|{\rm d}\rangle$
as discussed above.
Due to the p-h asymmetry of the underlying model these two singlet states will
have very different energies, such that we can discard the high-energy state
in the low-energy theory.
With this we directly arrive at an infinite-$U$ pseudogap Anderson model.

We conclude that the phase transitions of the pseudogap Anderson and
Kondo models are in the same universality classes -- this is supported
by the numerical calculations of Ref.~\onlinecite{GBI}.
For small $r$ the phase transition is naturally described in the Kondo
language of Sec.~\ref{sec:weak}, implying that the critical fixed point
of the Anderson model is located in the Kondo limit.
In contrast, for larger $r$ the formulation in terms of the Anderson model
provides the relevant degrees of freedom, in other words, both spinful and
spinless (many-body) impurity states play a role in the critical dynamics.


\section{Finite magnetic field}
\label{sec:field}

Interesting physics obtains in the pseudogap Kondo and Anderson models
in the presence of a finite magnetic field.
We concentrate here on the effect of a {\em local} field, applied to the
impurity only.
Note that a finite field applied to the bulk can modify the low-energy
behavior of the bath DOS due to Zeeman splitting;
then, pseudogap Kondo physics survives only for energy scales above
the Zeeman energy.

\begin{figure}[t]
\epsfxsize=2.7in
\centerline{\epsffile{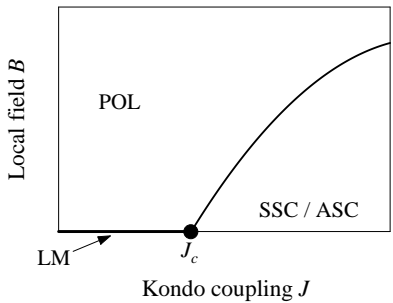}}
\caption{
Schematic $T=0$ phase diagram of the pseudogap Kondo model in a local
magnetic field $B$.
At $B=0$, there is a phase transition at $\Jb=J_c$ between LM and a strong-coupling phase,
SSC or ASC.
Upon application of a field, an unscreened spin becomes strongly polarized (POL),
whereas a screened spin in a strong-coupling phase is only weakly polarized.
Increasing the field at $\Jb>J_c$ drives a phase transition -- such a transition
is {\em not} present in the metallic case, $r=0$.
The phase diagram for a pseudogap Anderson model is similar.
As discussed in the text, the zero-field critical fixed point is always unstable w.r.t.
finite $B$ in the RG sense; thus the two transitions are in general in different
universality classes (which is already clear from symmetry considerations).
}
\label{fig:field}
\end{figure}

For $T=0$ and a metallic density of states, $r=0$, a local field $B$ only leads
to a crossover as function of $B/T_{\rm K}$, and all observables evolve smoothly from
$B=0$ (screened spin) to large $B$ (polarized spin).
In contrast, for any $r>0$ and $t>0$ (i.e. $\Jb>J_c$ in the Kondo model)
there will be a zero-temperature phase transition as function of the local field
between one phase with weak and one with strong impurity spin polarization,
see Fig.~\ref{fig:field}.
We briefly discuss this phase transition in the following -- it turns
out that the variables of the Anderson model provide a natural language
to analyze the problem.
Remarkably, all ingredients needed for the discussion of the
critical properties have already been calculated in the previous sections.

Strategically, we first discuss the decoupled impurity, and then analyze
the modifications arising from the presence of the hybridization term.
For the decoupled impurity, the local field is trivially a relevant
perturbation, with scaling dimension unity.
Thus, in the low-energy limit the field is effectively infinite, and
one of the two impurity states $|\uparrow\rangle$, $|\downarrow\rangle$
can be discarded (we will discard $|\downarrow\rangle$ in what follows).

\subsection{Asymmetric Anderson model}

In the p-h asymmetric case two impurity states have to be considered,
namely $|\uparrow\rangle$ and $|{\rm e}\rangle$.
We discuss a level crossing transition of these two, tunable e.g. by varying the
magnetic field, and being coupled to conduction electrons.
At the transition we are left with a (spinless) resonant level model,
and the analysis for a pseudogap host density of states is
in Sec.~\ref{sec:rlv}.
In particular, for $r>1$ the phase transition is a level crossing
with perturbative corrections, and for all $0<r<1$ we have a continuous
transition with a critical fixed point identical to the intermediate-coupling
fixed point of the resonant level model, Eq.~(\ref{g_rlv}).
Thus, the properties of the transition evolve smoothly as function of $r$ for
$0<r<1$, in contrast to the zero-field situation (!).

\subsection{Symmetric Anderson model}

In the presence of p-h symmetry and magnetic field, the decoupled impurity has
three low-energy states: $|\uparrow\rangle$, $|{\rm e}\rangle$, $|{\rm d}\rangle$.
The resulting level-crossing transition is technically identical to
the one of the zero-field infinite-$U$ Anderson model of Sec.~\ref{sec:infu}:
the two situations can be mapped onto each other via the p-h
transformation (\ref{ph2}).
From Sec.~\ref{sec:infu} we can read off the properties of the field-tuned
phase transition:
For $r>1$ we have again a level crossing with perturbative corrections.
A RG expansion around the level-crossing fixed point can be used
to calculate the critical properties below $r=1$.
This expansion describes the physics for all $r>r^\ast$.

It is interesting to ask what happens for $r<r^\ast$.
At $r^\ast$ the broken level symmetry is dynamically restored at the critical
fixed point, meaning that for $r<r^\ast$ the transition is controlled by a
zero-field critical fixed point.
From Secs.~\ref{sec:weak} and ~\ref{sec:sym} we know that for $r<\frac{1}{2}$
such an Anderson model transition with four degenerate impurity states is
the same as described by the weak-coupling RG in the Kondo language.
However, the mapping (\ref{ph2}) shows that we have to consider a Kondo model
of a charge pseudospin here, i.e., the corresponding Schrieffer-Wolff transformation
will project out the $|\uparrow\rangle$ and $|\downarrow\rangle$ states.
In other words, in a p-h symmetric situation with $0<r<r^\ast$,
where the zero-field transition is controlled by the SCR fixed point,
the finite-field transition will be asymptotically controlled by the
SCR' fixed point!

\subsection{Kondo model}

The above statement is consistent with a simple generalization of the
weak-coupling RG of Sec.~\ref{sec:weak} to the case of finite field.
One easily finds that the field is a relevant perturbation at the SCR
fixed point with a scaling dimension of unity, and the finite-field transition
is {\em not} accessible in a description using the Kondo model spin
variables.
(In contrast, the magnetic field is irrelevant at SCR'.)

\subsection{Symmetries and pseudospin field}

In the Anderson model language, the local field lifts the degeneracy
of the magnetic doublet $|\uparrow\rangle$, $|\downarrow\rangle$.
It breaks the SU(2) symmetry in the spin sector, and the
effect is clearly independent of the field direction.

As discussed in Sec.~\ref{sec:models}, the p-h symmetric Anderson model
display SU(2) symmetry also in the charge sector (charge pseudospin).
This can be broken by choosing $\Ub\neq -2\epsfb$, corresponding to a
pseudospin field in $z$ direction.
Clearly, a pseudospin field in perpendicular direction will have the same
effect -- this field corresponds to a local pairing field coupling to
$f_\uparrow f_\downarrow$.
Thus, the SU(2) pseudospin symmetry
shows that the effect of $s$-wave pairing correlations on the Kondo problem
is similar to a particle-hole asymmetry, as argued in Ref.~\onlinecite{sakaisc}
using a completely different approach.


\section{Conclusions}

In this paper, we have analyzed the zero-temperature phase transitions
in the pseudogap Kondo problem, characterized by a bath density of states
$\rho(\omega)\propto|\omega|^r$.
Using various perturbative RG expansions, formulated in the degrees
of freedom of either the Kondo or the Anderson impurity model,
we have developed critical theories for the phase transitions
in both the p-h symmetric and asymmetric cases.
We have established that $r=0$ and $r=\frac 1 2$ play the role of two lower-critical
dimensions in the p-h symmetric case, i.e., the non-trivial phase transition
disappears both as $r\to 0^+$ and $r\to\frac 1 2^-$ with diverging
correlation length exponent.
In contrast, $r=1$ is an upper-critical dimension for the p-h asymmetric
model.
The transitions for $0<r<1$ are described by interacting field theories
with {\em universal} local-moment fluctuations and
strong hyperscaling properties including $\omega/T$ scaling of dynamical
quantities.
In contrast, for $r\!>\!1$ we find a level crossing with perturbative
corrections, and hyperscaling is violated.

In the p-h symmetric case, we found two different expansions, one
around $r=0$ and one around $r=\frac{1}{2}$, to describe the same
critical fixed point.
In the presence of p-h asymmetry, a different critical fixed point
emerges, which can be analyzed in an expansion around $r=1$.

Apart from the small-$r$ expansion of Sec.~\ref{sec:weak},
all our theories were formulated using the Anderson model language.
This shows that the quantum phase transition between
a screened and an unscreened moment in Kondo-type models
can be nicely interpreted by saying that the system fluctuates
between ``possessing a moment'' and ``possessing no moment'' -- this is
precisely what is described by the effective Anderson model
at its valence-fluctuation fixed point.

We have calculated a number of observables, with results being in
excellent agreement with numerical data.
In particular, we have found an exact exponent for the conduction
electron $T$ matrix, valid for all expansions used in this paper.
We have also discussed the physics of the pseudogap Kondo problem
in a local magnetic field, where we have shown that -- in contrast to
the metallic Kondo effect -- a sharp transition occurs as a function
of the field.

Applications of our results include impurity moments in unconventional
superconductors \cite{bobroff1,bobroff2,MVRB,tolya} and other pseudogap
systems, like e.g. graphite.
On the theoretical side, we expect that the analysis of quantum
impurity models using Anderson instead of Kondo model variables may be
useful in a variety of problems.
Thus, field theories similar to ours can possibly be constructed for
other impurity quantum transitions, and will also be useful for
the study of lattice models in dynamical mean-field theory \cite{dmft} and
its extensions \cite{edmft}, where local quantum criticality can
be captured using effective single-impurity models.


\acknowledgments

We thank L. Balents, R. Bulla, M. P. A. Fisher, S. Florens, M. Kir\'{c}an,
K. Ingersent, N. Read, A. Rosch, Q. Si, P. W\"olfle and in particular
S. Sachdev for illuminating discussions,
and R. Bulla and T. Pruschke for help with the NRG calculations.
This research was supported by the Deutsche Forschungsgemeinschaft
through the Center for Functional Nano\-structures Karlsruhe.
MV also acknowledges support by the National Science Foundation
under Grant No. PHY99-0794 and the hospitality of KITP Santa Barbara
where part of this work was carried out.


\appendix

\section{The non-interacting resonant level model}
\label{app:rl}

Here we provide a few details about the non-interacting resonant level
model with a pseudogap density of states. The model is exactly solvable,
and all properties can be directly evaluated using the dressed
$f$ electron propagator (\ref{dressedf}).
The local susceptibility has been quoted in Sec.~\ref{sec:rlv}.
The $T$ matrix is given by $T(\omega) = \hybb^2 G_f(\omega)$;
a brief discussion of spectral properties can also be found in the Appendix
of Ref.~\onlinecite{GBI}.

\begin{figure}[t]
\epsfxsize=2.8in
\centerline{\epsffile{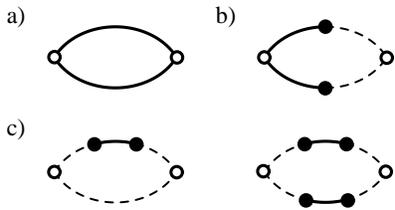}}
\caption{
Feynman diagrams for the spin susceptibility of the
non-interacting resonant level model of Sec.~\protect\ref{sec:rlv}.
Dashed/full lines denote $c_\sigma$/{\em dressed} $f_\sigma$ propagators,
respectively; the full dots are $V_0$ vertices, and the open circles are
sources.
a) $\chi_{\rm imp,imp}$.
b) $\chi_{\rm u,imp}$.
c) $\chi_{\rm u,u}$.
Note that the displayed diagrams give the {\em exact} results for the
susceptibilities, as all self-energies are contained in the dressed $f$
propagator.
}
\label{fig:rlvdgr1}
\end{figure}

We now sketch the calculation of $T\chi_{\rm imp}$ and $S_{\rm imp}$.
The diagrams contributing to $T\chi_{\rm imp}$ are in Fig.~\ref{fig:rlvdgr1};
note that these contain full $f$ propagators, and no higher-order
diagrams appear.
Power counting shows that terms of $1/T$ form do only arise from $\chi_{\rm{u,u}}$;
both $\chi_{\rm{imp},\rm{imp}}$ and $\chi_{\rm{u,imp}}$ are less singular.
The evaluation of the diagrams in Fig.~\ref{fig:rlvdgr1}c gives
\begin{eqnarray}
\chi_{\rm{u,u}}(T) &=&
\frac{V_0^2}{A_0} \, T\sum_n
\int_{-\Lambda}^\Lambda dk |k|^r \frac{|\omega_n|^{-r}}{(i\omega_n-k)^3} \\
&+&
\frac{V_0^4}{2 A_0^2} \, T\sum_n \left(
\int_{-\Lambda}^\Lambda dk |k|^r  \frac{|\omega_n|^{-r}}{(i\omega_n-k)^2} \right)^2 .
\nonumber
\end{eqnarray}
All factors of $V_0$ cancel exactly against the $A_0$, and
we obtain a quantity with a universal prefactor.
The $k$ integrals can be performed in the limit of infinite UV cutoff,
followed by the Matsubara summations.
The result is of Curie form, with
\begin{equation}
\chi_{\rm{u,u}}(T) = \frac{r}{8T} \,.
\end{equation}
Note that we have used the dressed $f$ propagators (\ref{dressedf}) in their
low-energy form -- due to their singular nature their high-energy properties
are unimportant for the leading low-temperature behavior of the
susceptibility.

The impurity entropy $S_{\rm imp}$ is directly calculated from the
full $f$ propagator (\ref{dressedf}).
Performing the tr ln in the partition function gives
\begin{equation}
\Omega_{\rm imp} = -2T \sum_n \ln [i{\rm sgn}(\omega_n)|\omega_n|^r] \,.
\end{equation}
One can easily see that the $T=0$ entropy contributions of $\ln(i\omega_n)$ and
$\ln|i\omega_n|$ are identical, thus we have
\begin{equation}
S_{\rm imp} = - \lim_{T\to 0}
\partial_T \left( -2 r \, T \sum_n \ln(i\omega_n) e^{i\omega_n 0^+} \right) \,.
\end{equation}
This is nothing but $2r$ times the entropy of a free spinless fermion,
resulting in $S_{\rm imp} = 2r\ln 2$.

\begin{figure}[t]
\epsfxsize=3.2in
\centerline{\epsffile{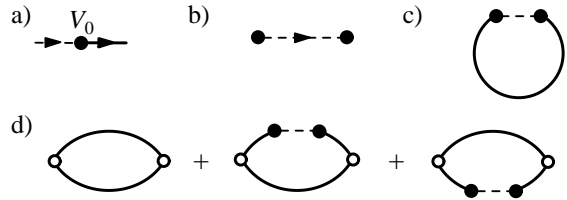}}
\caption{
Diagrams occurring in the perturbative RG for the
non-interacting resonant level model of Sec.~\protect\ref{sec:rlv}.
Here, full lines are {\em bare} $f_\sigma$ propagators.
a) Bare interaction vertex $\hybb$.
b) $f_\sigma$ self-energy.
c) Diagram for the $\hyb^2$ entropy correction.
d) Diagrams entering the local susceptibility to order $\hyb^2$.
Within the renormalized perturbation theory scheme all
corrections beyond second order in $\hyb$ vanish.
}
\label{fig:rlvdgr2}
\end{figure}

Interestingly, the above exact universal results can also be obtained
in the framework of a RG expansion in the hybridization, i.e., in a formal expansion
around $r=1$.
Here one exploits the intermediate-coupling nature of the fixed point, by
calculating observables using renormalized perturbation theory and employing
the fixed-point value (\ref{g_rlv}) of the coupling $\hyb$.
The necessary diagrams, Fig.~\ref{fig:rlvdgr2}, now contain {\em bare}
(instead of dressed) $f$ propagators.
Technically, higher-order terms appear in this expansion, but
upon interpreting the lowest-order result of renormalized perturbation theory
(e.g. re-exponentiating logarithms) these terms are completely
summed up.
Thus, within the RG framework the information about the non-trivial $f$ scaling dimension
is contained in the coupling value $\hyb^\ast$ instead of in the propagator $G_f$.
We briefly demonstrate the idea by evaluating the impurity entropy.
As we expand around $\hyb=0$ (the FImp fixed point) we obtain a perturbative correction
to the impurity thermodynamic potential which is of order $\hybb^2$
(Fig.~\ref{fig:rlvdgr2}b).
Taking the temperature derivative leads to
\begin{eqnarray}
\Delta S_{\rm imp} =
- \frac{\hybb^2}{2} T^{-2\bar{r}}
\int_{-\Lambda/T}^{\Lambda/T} d x |x|^r \cosh^{-2} \frac{x}{2}
\end{eqnarray}
with $\bar{r} = (1-r)/2$ as above.
Performing the integral in the limits $\Lambda\to\infty$ and $r\to 1$
and introducing the renormalized hybridization $\hyb$ yields
\begin{equation}
S_{\rm imp} = (1 - 2 \hyb^2) \, \ln 4 \,.
\end{equation}
With the fixed point value of the hybridization we obtain the
result $S_{\rm imp} = 2r\ln 2$ as above.

\section{RG for the interacting resonant-level model}
\label{app:rgsym}

In this appendix we present details of the renormalization group treatment
for the symmetric Anderson model in the vicinity of $r=\frac{1}{2}$.
It is based on an expansion around the SSC fixed point, i.e.,
around a non-interacting resonant-level model, with the interaction
strength being the expansion parameter.
Thus the expansion is done around an intermediate-coupling fixed point (!).

The starting point is the action (\ref{symact}) derived in
Sec.~\ref{sec:SSCRG}, for $0<r<1$.
Importantly, the ``bare'' (i.e. $\Ub=0$) $f$ propagator in (\ref{symact})
behaves as $1/(A_0\omega^r)$ at low energies, Eq.~(\ref{dressedf}),
as the conduction electrons have already been integrated out.
The prefactor $A_0$ will be kept explicitly in the renormalization
steps.
We employ the field-theoretic RG scheme in order to determine the flow
of the coupling renormalized coupling $\U$, introduced in Eq. (\ref{udef}),
with dimensional regularization and minimal subtraction of poles.
As will be seen below, the lowest non-trivial renormalizations arise at
two-loop order.

\begin{figure}[t]
\epsfxsize=3.3in
\centerline{\epsffile{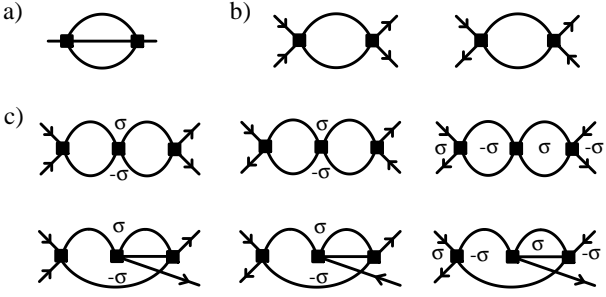}}
\caption{
Diagrams for the RG of the symmetric Anderson impurity model, with notation
as in Fig.~\protect\ref{fig:symdgr1}.
a) $\Ub^2$ self-energy contribution, which vanishes.
b) One-loop vertex renormalizations, which vanish due to p-h symmetry.
c) Two-loop vertex renormalizations -- these are the only contributions
to the beta function (\protect\ref{betau}).
}
\label{fig:symdgr2}
\end{figure}

The needed diagrams arising in the RG treatment
are displayed in Fig.~\ref{fig:symdgr2}.
A number of observations can be made:
(i)  Hartree diagrams vanish due to the overall p-h symmetry of the model.
(ii) Explicit calculation shows that the one-loop vertex renormalization diagrams,
i.e., the $\U^2$ corrections to $\U$ in Fig.~\ref{fig:symdgr2}b,
do not develop $\epsilon$ poles.
This also follows from the fact that the beta function cannot contain even powers
in $\U$ due to p-h symmetry.
(iii) Explicit evaluation shows that the $f$ self-energy up to two-loop
order, Fig.~\ref{fig:symdgr2}a,
does not contribute singular propagator renormalizations.
In other words, the field renormalization factor is
\begin{equation}
Z=1
\end{equation}
to two-loop order.
As we will argue below, this result is {\em exact} to all orders in
perturbation theory.
It implies that the $f$ field does not acquire an anomalous
dimension, $\eta_f=0$.

The two-loop diagrams for the renormalization of $\U$ can be divided into two
groups of three each, displayed in the two lines of Figs.~\ref{fig:symdgr2}c.
The diagrams in each group can be easily seen to be equal.
Explicitly:
\begin{eqnarray}
\label{rgs1}
{\rm (13c1)} =
\Ub^3 \left[ \int \frac{d\omega_1}{2\pi} \, G_f(i\omega_1) G_f(i\omega_1+i\nu) \right]^2
\end{eqnarray}
where $\nu$ is the sum of two external frequencies.
Using the explicit form of $G_f$ (\ref{dressedf}), the frequency integral can be
split according to the absolute value in $G_f$, and then performed directly
with UV cutoff sent to infinity.
The result contains $\epsilon$ poles and reads
\begin{eqnarray}
{\rm (13c1)} = \Ub^3 \frac{\nu^{2-4r}}{\pi^2}
\left[\frac{1}{\epsilon^2} + \frac{\pi - 2\ln 4}{\epsilon} + {\cal O}(\epsilon^0) \right] .
\end{eqnarray}
The second group of diagrams in Fig.~\ref{fig:symdgr2}c is slightly
more complicated. All three of them can be brought into the form
\begin{eqnarray}
{\rm (13c4)}
&=& 2 \Ub^3 \int \frac{d\omega_1}{2\pi} \frac{d\omega_2}{2\pi}\,
G_f(i\omega_1) G_f(\omega_2) \\
&& ~~~~\times\, G_f(-i\omega_1+i\nu) G_f(i\omega_2-i\omega_1+i\omega)
\nonumber
\end{eqnarray}
where the prefactor 2 accounts for the different arrangement of the vertices,
and $\omega$ is a (third) external frequency.
The $\omega_2$ integral can be performed directly and leads to the same expression
that occurred in the brackets in (\ref{rgs1}).
The remaining $\omega_1$ integral can also be performed after splitting the
interval into four parts.
Straightforward algebra,
using the identity $\arcsin\sqrt{x} - i {\rm arctanh}\sqrt{1-1/x} = \pi/2$,
where $x=\omega/\nu < 1$,
leads to
\begin{eqnarray}
{\rm (13c4)}
= \Ub^3 \frac{\nu^{2-4r}}{\pi^2}
\left[-\frac{1}{\epsilon^2} - \frac{3(\pi\!-\!2\ln 4)}{2\epsilon} + {\cal O}(\epsilon^0) \right].
\end{eqnarray}
Upon adding the contributions from \ref{fig:symdgr2}c1--\ref{fig:symdgr2}c6,
only single poles in $\epsilon$ remain -- as $Z=1$, the cancellation of the double
poles is a consistency check of our calculation.
Minimal subtraction of poles at external frequencies set to $\mu$
yields the renormalization factor for the quartic coupling:
\begin{eqnarray}
Z_4 = 1 + \frac{3(\pi-2\ln 4)\U^2}{2\pi^2\epsilon} \,.
\end{eqnarray}
The RG beta function can be evaluated by taking the $\mu$
derivatives of (\ref{udef}) at fixed bare coupling,
\begin{eqnarray}
\beta(\U) \equiv \mu \frac{d \U}{d \mu} \bigg|_{\Ub}
=
\epsilon \U \left(1 -  \frac{3(\pi-2\ln 4)}{\pi^2} \frac{\U^2}{\epsilon}\right) \,,
\end{eqnarray}
which is the result in Eq.~(\ref{betau}).

We now provide the proof for $Z=1$, i.e., no singular propagator renormalizations
occur in the present problem. What is needed is p-h symmetry and the form of
the bare propagator, $G_f\propto \omega^{-1/2}$ or $G_f(\tau)\propto\tau^{-1/2}$ in the
low-energy limit for $r=\frac 1 2$.
We argue that no logarithms arise when evaluating the self-energy at $r=\frac 1 2$.
By power counting, all $f$ self-energy diagrams in the $\tau$ domain give a
result proportional to $\tau^{-3/2}$.
Let us discuss internal $\tau$ integrals arising in the evaluation (which could produce logs).
In the first step, those involve four $G_f$ propagators due to the quartic interaction --
here it is important that tadpoles do not contribute due to p-h symmetry.
Therefore, the integrand will behave as $\tau^{-2}$, and logs cannot occur.
In further steps, four propagators do not necessarily occur, but the power counting
argument shows that each internal time still has to occur with a power of $\tau^{-2}$.
This shows that the integrands will always be {\em more singular} than $1/\tau$,
and no logarithms arise.
Clearly, such an argument cannot be constructed for the vertex diagrams, as they
behave as $1/\tau$ at $r=\frac 1 2$, and Fourier transformation generically yields
an $\epsilon$ pole.

Let us turn to the local susceptibility renormalization factor $Z_\chi$.
As described in Sec.~\ref{sec:symobs}, one can either determine
the composite operator renormalization factor $Z_2$ and use $Z_\chi=Z_2^2$,
or evaluate $Z_\chi$ directly with the help of the perturbative
correction to $\chi_{\rm loc}$ in Fig.~\ref{fig:symdgr1}c.
In both cases, the $f$ bubble diagram is needed.
To leading order in $\epsilon$ it evaluates to
\begin{eqnarray}
T\sum_n G_f(i\omega_n) G_f(i\omega_n+i\nu) = \frac{|\nu|^\epsilon}{\pi\epsilon}
\end{eqnarray}
in the zero-temperature limit.
For the correlation function $G_f^{(2,1)}$ in Fig.~\ref{fig:symdgr1}b
we introduce a renormalization factor as follows:
$G_f^{(2,1)} = Z_2 Z^{-1} G_{f,R}^{(2,1)}$.
Demanding that the renormalized $G_{f,R}^{(2,1)}$ is free of poles
at scale $\mu$, and using $Z=1$ we find:
\begin{equation}
Z_2 = 1 + \frac{\U}{\pi\epsilon} \,,
\end{equation}
which gives $Z_\chi$ in Eq.~(\ref{zchisym}).


\section{RG for the infinite-$U$ Anderson model}
\label{app:rginfu}

The infinite-$U$ Anderson model can be analyzed in the vicinity of
$r=1$ using an expansion in the hybridization strength.
Here we describe details of the field-theoretic RG, with
the results appearing in Sec.~\ref{sec:infu}.
Starting point is the action (\ref{th}), written in terms of
auxiliary fields $f_\sigma$ and $b_s$.
The RG proceeds by perturbatively calculating renormalizations
to the propagators and vertices appearing in (\ref{th}),
which yields expression for the renormalization factors
defined in Eq.~(\ref{gdef}).

The relevant diagrams are displayed in Fig.~\ref{fig:infudgr2}.
Note that there is no one-loop contribution to the vertex renormalization.
At one-loop order we have the graphs for the $f$ and $b$ self-energies.
Evaluation of the first diagram for the $f$ self-energy in
Fig.~\ref{fig:infudgr2}a gives:
\begin{eqnarray}
{\rm (14a1)} = \hybb^2 \, T \sum_{i\omega_n}
\int \frac{d k\,|k|^r}{i\omega_n\!-\!k} \,
\frac{1}{i\nu \!-\! i\omega_n \!-\!\lam} \,.
\end{eqnarray}
We work at criticality, i.e., $\epsfb=0$.
First the frequency sum can be performed, here the $\lam\to\infty$ limit
is important: it suppresses contributions from positive $k$.
The remaining $k$ integral is UV convergent, and we finally obtain
\begin{eqnarray}
{\rm (14a1)} &=& - \frac{\hybb^2}{2\bar{r}} \, (i\nu - \lam)^r \nonumber\\
       &=& - A_\mu \frac{\hyb^2}{2\bar{r}} (i\nu - \lam) \,.
\end{eqnarray}
In the second line we have expressed the result in terms of renormalized quantities,
with $A_\mu = \mu^{2\bar{r}} (i\nu - \lam)^{-2\bar{r}} Z_\hyb^2/(Z_f Z_b)$.
Demanding cancellation of poles in the expressions for the renormalized
$f$ Green's function at external frequency $i\nu - \lam = \mu$
we obtain the one-loop result for the $f$ renormalization factor
as
\begin{equation}
Z_f = 1 - \frac{\hyb^2}{2\bar{r}} \,.
\end{equation}
The other diagrams in Fig.~\ref{fig:infudgr2}a,b,c are evaluated in a similar
manner.
In the two-loop self-energies, the real part of the inner self-energy insertions,
which diverges linearly with the UV cutoff, is exactly cancelled by the
corresponding counter-term.

\begin{figure}[!t]
\epsfxsize=3.2in
\centerline{\epsffile{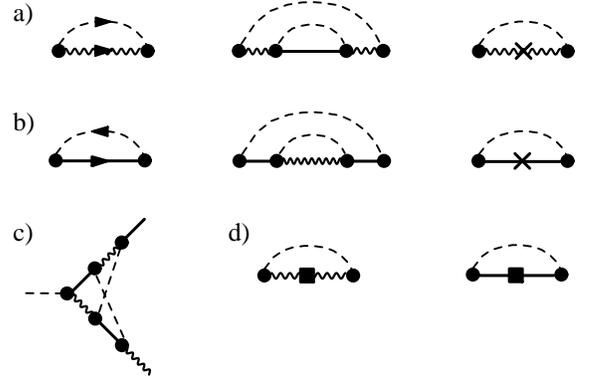}}
\caption{
Diagrams for the RG of the infinite-$U$ Anderson model.
Notations are as in Fig.~\protect\ref{fig:infudgr1}.
a),b) Self-energies for $f_\sigma$ and $b_s$ to two-loop order.
The crosses denotes the counter-terms which cancel the real parts of the
self-energy insertions.
c) Two-loop vertex renormalization; there is no contribution to one-loop order.
d) Diagrams with mass insertions (squares), needed to determine the correlation
length exponent.
}
\label{fig:infudgr2}
\end{figure}

Collecting all expressions yields the renormalization factors,
defined in Eq.~(\ref{gdef}), to two-loop order as
\begin{eqnarray}
Z_f &=& 1 - \frac{\hyb^2}{2\bar{r}} - \left(
            \frac{1}{4\bar{r}^2} + \frac{1}{2\bar{r}}\right)\hyb^4 \,, \nonumber \\*
Z_b &=& 1 - \frac{\hyb^2}{\bar{r}} - \left(
            \frac{1}{4\bar{r}^2} + \frac{1}{2\bar{r}}\right)\hyb^4 \,, \nonumber \\*
Z_\hyb &=& 1 + \frac{1}{4\bar{r}} \hyb^4
\,.
\end{eqnarray}
With these results, we can take the logarithmic $\mu$ derivative of
Eq.~(\ref{gdef}) at fixed bare coupling, we obtain the beta function
\begin{eqnarray}
\beta(\hyb) \equiv \mu \frac{d \hyb}{d \mu} \bigg|_{\hybb}
\end{eqnarray}
as quoted in Eq.~(\ref{betaginfu}).
The anomalous field dimensions are obtained from
\begin{eqnarray}
\eta_f &=& \mu \frac{d \ln Z_f}{d \mu} = \beta(\hyb) \frac{d\ln Z_f}{d\hyb} =   \hyb^2 + 2 \hyb^4\\*
\eta_b &=& \mu \frac{d \ln Z_b}{d \mu} = \beta(\hyb) \frac{d\ln Z_b}{d\hyb} = 2 \hyb^2 + 2 \hyb^4\,.
\end{eqnarray}

To determine the flow away from criticality and the correlation length
exponent, we follow the standard scheme \cite{bgz} via insertions of composite
operators, representing mass terms.
Physically, only the difference between the masses of the $f$ and $b$ auxiliary
fields is relevant, and introducing one type of composite operator is sufficient.
In the following we work with $\bar{f}f$ insertions, which acquire a corresponding
renormalization factor $Z_{2f}$.
To determine $Z_{2f}$ we consider a correlation function
$G_{bf}^{(2,1)} = \langle\langle b^\dagger(\tau) b(\tau'); (f^\dagger f)(\tau'') \rangle\rangle$.
The renormalization factor $Z_{2f}$ is then defined through
\begin{equation}
G_{bf}^{(2,1)} = \frac{Z_{2f}}{Z_f} Z_b G_{bf,R}^{(2,1)}
\end{equation}
Evaluating the diagram in Fig.~\ref{fig:infudgr2}d and demanding cancellation of
poles in $G_{bf,R}^{(2,1)}$ gives, to one-loop accuracy,
\begin{equation}
Z_{2f} = 1 + \frac{\hyb^2}{\bar{r}}.
\end{equation}
The correlation length exponent can be related to the $Z$ factors via the RG equation \cite{bgz} for
the renormalized $G_{bf}^{(2,1)}$, according to
\begin{eqnarray}
\frac{1}{\nu}-1 = \mu \frac{d}{d \mu} \ln\frac{Z_{2f}}{Z_f}
\end{eqnarray}
which yields
\begin{equation}
\frac{1}{\nu} = 1 - 3 {\hyb^\ast}^2 \,.
\end{equation}
The result for $\nu$ is in Eq.~(\ref{nuz_infu}), and is of course identical to the
one obtained in Ref.~\onlinecite{MVLF} using the familiar momentum-shell method.

The local susceptibility exponent is determined via a renormalization factor, $Z_\chi$,
for the two-point correlations of the impurity spin.
The leading diagrams are in Fig.~\ref{fig:infudgr1}c;
the $\hyb^2$ diagrams are identical, with a leading singular contribution of
\begin{eqnarray*}
&&\frac{1}{2} \hybb^2
\int_0^\tau d\tau_2 \int_0^{\tau_2} \! d\tau_1
G_{c0}(\tau_2-\tau_1) = \frac {\hybb^2} {4\bar{r}} \tau^{2\bar{r}}
\end{eqnarray*}
each. Note that the counter-term contributions can be ignored here,
as they do not develop poles in $\bar{r}$.
Demanding cancellation of poles yields
\begin{equation}
Z_\chi = 1- \frac{\hyb^2}{\bar{r}}.
\label{zchi3}
\end{equation}
Taking the logarithmic derivative w.r.t. $\mu$ one obtains the
$\eta_\chi$ value quoted in Sec.~\ref{sec:infuobs}.

The renormalization factor for the $T$ matrix (diagrams not shown)
is obtained as
\begin{equation}
Z_T = 1 - \frac{3}{2} \frac{\hyb^2}{\bar{r}} - \frac 3 2 \frac{\hyb^4}{\bar{r}}
\end{equation}
to two-loop order.
The resulting anomalous exponent $\eta_T$ fulfills the exact equation $\eta_T= 2\bar{r}$.


\section{Spin anisotropies}
\label{app:aniso}

Thus far our discussion has been restricted to impurity models with full SU(2)
spin symmetry.
In this appendix we show that spin anisotropies, e.g. in the Kondo interaction,
are irrelevant at all critical points considered.

For the weak-coupling RG of Sec.~\ref{sec:rgweak}, formulated in the variables of
the Kondo model, we introduce Kondo couplings $J_\perp$ and $J_z$ for the
transverse and longitudinal part of the Kondo interaction.
The RG equation (\ref{betaj}), to one-loop order, generalizes to
\begin{eqnarray}
\beta(j_\perp) &=& r j_\perp - j_\perp j_z \,, \nonumber \\
\beta(j_z)     &=& r j_z - j_\perp^2 \,.
\end{eqnarray}
For the metallic case, $r=0$, there is a line of fixed points
at $j_\perp=0$, $j_z<0$, representing an unscreened moment.
In contrast, for $r>0$ only SU(2) symmetric fixed points survive,
namely LM with $j_\perp^\ast=j_z^\ast=0$ and SCR with
$j_\perp^\ast=j_z^\ast=r$.
Thus, the symmetric critical fixed point of the Anderson and
Kondo models is stable w.r.t. SU(2) symmetry breaking.

Turning to the p-h asymmetric model, we start from the infinite-$U$
Anderson model (\ref{aiminfu}) and introduce a spin dependence
in the hybrization, ${\hybb}_\uparrow \neq {\hybb}_\downarrow$.
As is easily seen via Schrieffer-Wolff transformation,
for a generic p-h asymmetric conduction band this is equivalent to
an anisotropic exchange interaction plus a local magnetic field.
Without loss of generality we restrict the following analysis to a p-h
symmetric conduction band, where the field term is absent.
The RG equation (\ref{betaginfu}), to one-loop order, generalizes to
\begin{eqnarray}
\beta(\hyb_\uparrow) &=&
- {\bar r} \hyb_\uparrow + \frac{\hyb_\uparrow}{2}(2\hyb_\uparrow^2 +  \hyb_\downarrow^2)
\,, \nonumber \\
\beta(\hyb_\downarrow) &=&
- {\bar r} \hyb_\downarrow + \frac{\hyb_\downarrow}{2}(2\hyb_\downarrow^2 + \hyb_\uparrow^2)
\,.
\end{eqnarray}
Apart from the SU(2) symmetric fixed point, ${\hyb_\sigma^2}^\ast = \frac{2}{3}\bar{r}$,
there are two other fixed points with
${\hyb_\uparrow^2}^\ast = \bar{r}$, $\hyb_\downarrow = 0$ and
${\hyb_\downarrow^2}^\ast = \bar{r}$, $\hyb_\uparrow = 0$ --
these are, however, infrared unstable w.r.t. finite $\hyb_\downarrow$ ($\hyb_\uparrow$).
Thus, the only stable critical fixed point is the SU(2) symmetric one, which corresponds
to the ACR fixed point analyzed in Sec.~\ref{sec:infu}.


\end{document}